\documentclass[a4paper,floatsintext,man,donotrepeattitle]{apa6}
\usepackage[utf8]{inputenc}
\usepackage[american]{babel}
\usepackage{csquotes}
\usepackage[style=apa,sortcites=true,sorting=nyt,backend=biber]{biblatex}
\DeclareLanguageMapping{american}{american-apa}
\bibliography{fddm.bib}

\usepackage{hyperref}
\usepackage{threeparttable}
\usepackage{booktabs}
\usepackage{amsmath, amsfonts}
\usepackage{graphicx}
\graphicspath{{./Images/}}
\usepackage[colorinlistoftodos]{todonotes}

\usepackage{mathtools,etoolbox}
\newcommand*\rfrac[2]{{}^{#1}\!/_{#2}}
\newcommand{\sett}[1]{\left\{#1\right\}}
\newcommand{\floor}[1]{\left\lfloor#1\right\rfloor}
\newcommand{\ceil}[1]{\left\lceil#1\right\rceil}
\newcommand{\abs}[1]{\left\lvert#1\right\rvert}
\DeclarePairedDelimiter{\angbrk}{\langle}{\rangle}
\newcommand{\Real}{\mathbb R}
\newcommand{\Int}{\mathbb Z}
\newcommand{\Nat}{\mathbb N}
\renewcommand{\th}{^{\text{th}}}
\usepackage{caption}
\usepackage{url} 
\usepackage{soul} 
\usepackage[toc,page]{appendix}
\usepackage{xcolor}
\usepackage{listings}  
\title{Another Approximation of the First-Passage Time Densities for the Ratcliff Diffusion Decision Model}
\shorttitle{fddm}
\twoauthors{Kendal Foster}{Henrik Singmann}
\twoaffiliations{University of Warwick}{University College London\\University of Warwick}
\authornote{Kendal Foster was supported by the Engineering and Physical Sciences Research Council and the Medical Research Council grant EP/L015374/1. Henrik Singmann was supported by SNSF grant 100014\_165591. We would like to thank Peter Strong for valuable comments.\\Corresponding author: Kendal Foster, EPSRC \& MRC Centre for Doctoral Training in Mathematics for Real-World Systems, Zeeman Building, University of Warwick, Coventry CV4 7AL, UK; email: \hbox{kendalfoster@gmail.com}\\The software developed for this paper, \texttt{fddm}, is available at: \url{https://cran.r-project.org/package=fddm}\\The source code of \texttt{fddm} as well as the code to recreate all results and figures is available at: \url{https://github.com/rtdists/fddm}}

\abstract{
	We present a novel method for approximating the probability density function (PDF) of the first-passage times in the Ratcliff diffusion decision model (DDM). We implemented this approximation method in \texttt{C++} using the \texttt{R} package \texttt{Rcpp} to utilize the faster \texttt{C++} language while maintaining the \texttt{R} language interface. In addition to our novel approximation method, we also compiled all known approximation methods for the DDM density function (with fixed and variable drift rate), including previously unused combinations of techniques found in the relevant literature. We ported these approximation methods to \texttt{C++} and optimized them to run in this new language. Given an acceptable error tolerance in the value of the PDF approximation, we benchmarked all of these approximation methods to compare their speed against each other and also against commonly used \texttt{R} functions from the literature. The results of these tests show that our novel approximation method is not only orders of magnitude faster than the current standards, but it is also faster than all of the other approximation methods available even after translation and optimization to the faster \texttt{C++} language. All of these approximation methods are bundled in the \texttt{fddm} package for the \texttt{R} statistical computing language; this package is available via CRAN, and the source code is available on GitHub.
}

\keywords{response times, evidence accumulation model, diffusion model, probability density function, software}

\begin{document}
\maketitle

\setcounter{secnumdepth}{3}

\newpage
\section{Introduction} \label{sec:int}

The Ratcliff \textit{diffusion decision model} (DDM) \parencite{ratcliff1978theory,ratcliff2008diffusion} is the most prominent evidence accumulation model for jointly modelling binary decision and associated response times. It assumes noisy information uptake; this noise is characterized by the Wiener process. The DDM is popular because (a) it utilizes a lot of information from the collected data, which in turn allows us to draw conclusions about the underlying cognitive processes that are assumed to underlie the empirical response time distribution \parencite{voss2008fast} and (b) it provides a good fit to data across different empirical domains \parencite[e.g.,][]{forstmann2016sequential}.

However, this efficient use of information comes at a rather high computational cost. The probability density function (PDF) of the DDM is the density of \textit{first-passage times}, that is, the time it takes for the Wiener process to first cross a fixed boundary. The computation of the PDF of the first-passage times is computationally expensive because it contains an infinite sum. Of course computers cannot actually evaluate an infinite sum, so several papers have developed approximation methods to this density function \parencite{voss2008fast, navarro2009fast, blurton2012fast, gondan2014even}. The different approximation methods provide different \textit{truncation} rules that guarantee the value of the approximated density function does not differ from its true value by more than a pre-specified criterion, $\epsilon$.

The analytic formulation of the diffusion model PDF originates from \textcite{feller1968introduction}, who provides its seminal derivation. In this derivation from first principles (i.e., the Wiener process), there is a step that requires taking a limit. \textcite{feller1968introduction} provides two different -- but equivalent -- limiting processes that yield two different -- but equal -- forms of the density function. Each of these forms contains an infinite sum, and they are known individually as the ``large-time'' density function and the ``small-time'' density function \parencite{navarro2009fast}. Even though the two distinct forms of the DDM density function are mathematically equivalent, they can produce slightly different results when calculated numerically with a fixed and limited precision.

The original truncations of the ``large-time'' and ``small-time'' density functions provided by \textcite{navarro2009fast} depend on \textit{precalculating} the number of terms for the \textit{truncated} versions of the aforementioned infinite sums. As suggested by their names, these density function approximation methods have contrasting performance across the domain of response times. The ``large-time'' approximation method is efficient for large response times, and the ``small-time'' approximation method is efficient for small response times. Contrarily, the ``large-time'' approximation method can be inefficient for small response times and vice-versa for the ``small-time'' approximation method; these inefficiencies may lead to inaccuracies in the approximations themselves. To make up for these shortcomings, \textcite{navarro2009fast} proposed a mechanism for choosing the more efficient and accurate approximation method between the ``large-time'' and ``small-time'' given a particular observation (i.e., pair of choice and associated response time) and set of parameter values. As shown in a set of benchmark analyses later in this manuscript, approximation methods that combine both the ``large-time'' and ``small-time'' density functions are not only faster on average but also more stable when used in an optimization setting.

\textcite{gondan2014even} introduced a new precalculation of the number of terms in the ``small-time'' truncated sum. Since this new precalculation performs the same role as that of \textcite{navarro2009fast}, we can implement it in the choosing mechanism provided by \textcite{navarro2009fast}. We include this combination of approximation methods in both our analysis and software package as it has not yet been implemented in existing software.

The main contribution of the present work is a novel approximation method to the ``small-time'' density function which -- in contrast to the existing approximation methods -- does not rely on precalculating the number of terms required in the truncated sum. By avoiding this precalculation, the new approximation methods relies on fewer overall computations when compared to the currently available approximation methods. We also provide a heuristic that allows the combination of the new ``small-time'' approximation method with the ``large-time'' approximation method of \textcite{navarro2009fast}. In a series of benchmark analyses, we show that the new combined approximation method is both faster than existing approximation methods and at the same time maintains the same high numerical accuracy, even when used for fitting real empirical data.

The existing literature \parencite{navarro2009fast,gondan2014even} discusses density function approximation methods to the variant of the DDM that maintains a constant drift rate across trials (i.e., the drift rate parameter $v$ is constant across experimental trials). However, it is common to allow the drift rate to vary across trials (i.e., include the parameter $\eta > 0$), and this produces slightly different density functions. More specifically, the inclusion of across-trial variability in the drift rate is one of the big contributions of \textcite{ratcliff1978theory} and is what allows the diffusion model to predict slow errors \parencite[e.g.,][]{donkin2018response}. Conveniently, the infinite sums in either version of the density function (``large-time'' or ``small-time'') is the same regardless of the inclusion of variability in the drift rate. We can exploit this sameness to apply the known approximation methods to the variable drift rate variants of the DDM density functions, being careful to ensure the desired error tolerance is scaled appropriately. As there is only one other implementation of a density function approximation method that includes variability in the drift rate \parencite{rtdists}, we include these variants both in our benchmark analyses and software package.

For both consistency and convenience in performing the benchmark analyses, we built the software package, \texttt{fddm}, that contains all of the available implementations of the DDM density function approximation methods. All of these implementations are available through the function \texttt{dfddm()} in the \texttt{fddm} package for the programming language \texttt{R} \parencite{R}. However, for computational efficiency, all approximation methods are implemented in pure \texttt{C++}, a relatively low-level programming language known for comparatively fast numerical computations. We used the \texttt{R} package \texttt{Rcpp} \parencite{Rcpp1, Rcpp2, Rcpp3}  to provide an \texttt{R} language user interface to the \texttt{C++} implementations. \texttt{fddm} is freely available from \texttt{CRAN}\footnote{\url{https://cran.r-project.org/package=fddm}} and comes with detailed documentation and model fitting examples.

Packaging all of the available implementations in a single software package is not only convenient for running benchmark analyses, but it also avoids any inconsistencies that may arise in the benchmark results from using implementations from different sources. The benchmark tests recorded the implementations' evaluation time given a predefined error tolerance, $\epsilon$, that remained consistent throughout the testing. Our analysis shows that the fastest and most stable approximation method for the DDM density functions of those currently available is our implementation of the ``large-time'' approximation method from \textcite{navarro2009fast} combined with the ``small-time'' approximation method formally introduced in this paper; we use this approximation method as the default for the \texttt{dfddm()} function. For reference, we also compare our implementations to existing DDM density function approximation implementations in \texttt{R} \parencite{rtdists, RWiener} and show that the default implementation in \texttt{dfddm()} is the fastest of those.

The remainder of this paper is structured as follows: Section \ref{sec:apx} will detail all of the approximations to the two infinite sums discussed in the literature. In particular, this section will highlight the differences across the density functions and their approximation methods. Additionally, we will present a novel method for approximating the ``small-time'' density function, and this method is faster than the comparable approximation methods from the literature given a standard accuracy. In Section \ref{sec:imp}, we will discuss how we designed the computer code for the implementations of the density function approximation methods; we translate existing code into \texttt{C++} and offer our optimizations for the existing approximation methods. The results of running benchmark tests on these implementations are discussed in Section \ref{sec:ben}, and we also include supplemental analyses to complement certain facets of the benchmark results in the Supplemental Materials document. Concluding remarks containing implications for how the field should use these approximation methods and implementations are found in Section \ref{sec:dis}. All additional mathematical content, such as proofs, is located in the Appendix at the end of this paper. Moreover, the source code for reproducing all analyses reported in the paper can be found on GitHub.\footnote{In the folder \texttt{paper\_analysis} of: \url{https://github.com/rtdists/fddm}}

\section{The DDM Density Functions} \label{sec:den}

As mentioned in the introduction, the density function of the DDM is the density of the first-passage times of the underlying Wiener process. The DDM density function is defined for data given as a tuple $\sett{c, t}$, where $c \in \sett{1, 2}$ is a binary choice and $t \in [0, \infty)$ is the associated response time. In the \texttt{fddm} software, we consider the DDM variant with six parameters: $v \in (-\infty, \infty)$, the drift rate; $\eta \in [0, \infty)$, the inter-trial variability in the drift rate\footnote{The parameter $\eta$ in this paper is named \texttt{sv} in the \texttt{fddm} software.}; $a \in (0, \infty)$, the threshold separation; $w \in (0, 1)$, the relative starting point in the diffusion process; $t_0 \in [0, \infty]$, the non-decision time; and $\sigma^2 \in [0, \infty)$, the diffusion coefficient of the underlying Wiener process. In the following subsections we will first explain our simplifications regarding the parameters for expositional purposes, and then we will explicitly provide the density functions using notation which closely resembles that of the recent literature \parencite{navarro2009fast, gondan2014even}.

\subsection{Our Interpretation of the Density Functions} \label{sec:den-our}

For clarity in exposition, we make some simplifications to the six-parameter variant of the DDM that we mentioned above. Although the \texttt{fddm} software includes all six of the aforementioned parameters, our subsequent discussions of the density functions and their properties will only detail the simplified versions that we will discuss imminently. Specifically, we exclude some of the inputs and parameters from our formulations because they are not critical to the mathematical understanding of the density functions. Ultimately, we will denote the density function as $f(t ~|~ v, \eta, a, w)$ by the end of this subsection.

\subsubsection{The Two Thresholds of the DDM} \label{sec:den-our-thr}

The DDM depends on the underlying Wiener process crossing one of two thresholds that are denoted by the binary choice $c \in \sett{1, 2}$ in the data tuple. As the density function of the DDM is the density of the Wiener first-passage times across these two thresholds, we define the full density of the DDM as $\text{WFPT}(c, t ~|~ v, \eta, a, w, t_0, \sigma^2)$. However, the PDF is actually calculated in a piecewise manner such that there is a different density function that corresponds to each threshold, $f_\text{lower}$ for $c = 1$ and $f_\text{upper}$ for $c = 2$, and this is described in the following equation:
\begin{equation} \label{eq:pdf_lu}
	\text{WFPT}(c, t ~|~ v, \eta, a, w, t_0, \sigma^2) =
	\begin{cases}
		f_\text{lower}(t ~|~ v, \eta, a, w, t_0, \sigma^2) & \text{ if $c = 1$,}\\
		f_\text{upper}(t ~|~ v, \eta, a, w, t_0, \sigma^2) & \text{ if $c = 2$.}
	\end{cases}
\end{equation}

Conveniently, there is a simple relationship between $f_\text{lower}$ and $f_\text{upper}$:
\begin{equation} \label{eq:lu}
	f_\text{upper}(t ~|~ v, \eta, a, w, t_0, \sigma^2) = f_\text{lower}(t ~|~ -v, \eta, a, 1-w, t_0, \sigma^2).
\end{equation}

Thus, we can write the full density function of the DDM as
\begin{equation} \label{eq:pdf_ll}
	\text{WFPT}(c, t ~|~ v, \eta, a, w, t_0, \sigma^2) =
	\begin{cases}
		f_\text{lower}(t ~|~ v, \eta, a, w, t_0, \sigma^2) & \text{ if $c = 1$,}\\
		f_\text{lower}(t ~|~ -v, \eta, a, 1-w, t_0, \sigma^2) & \text{ if $c = 2$.}
	\end{cases}
\end{equation}

Technically, $f_\text{lower}$ and $f_\text{upper}$ are defective density functions when considered individually as only WFPT integrates to unity, and $f_\text{lower}$ and $f_\text{upper}$ usually do not. We are interested, however, in calculating these defective density functions because they are more useful since it is straightforward to obtain the full PDF from either of these two pieces. Following the precedent set by the literature \parencite{navarro2009fast, gondan2014even}, we will work with the density function at the lower threshold and refer to this simply as either the \textit{density function} or $f$ for the remainder of this paper.

\subsubsection{The Underlying Wiener Process} \label{sec:den-our-wie}

As a reminder, the DDM provides the density function of the first-passage times of the Wiener process, a biased random walk in continuous time. This underlying Wiener process has a drift rate (i.e., the tendency toward one of the thresholds) and a diffusion coefficient (i.e., the amount of noise in the diffusion process). The diffusion coefficient is denoted $\sigma^2$ and only scales the other parameters (i.e., $v \to \frac{v}{\sigma}$, $\eta \to \frac{\eta}{\sigma}$, and $a \to \frac{a}{\sigma}$), so we fix $\sigma^2 = 1$ to both simplify our mathematical expressions and follow the precedent set by the recent literature \parencite{navarro2009fast, blurton2012fast, gondan2014even, blurton2017first}.\footnote{The default value in the \texttt{fddm} software is $\sigma^2 = 1$ as well.} Note that other formulations in the literature fix $\sigma^2 = 0.01$ (most notably \textcite{ratcliff1978theory}), so the necessary care must be taken to rescale the parameters when comparing the results of different formulations.

For each trial in our parameterization of the DDM density functions, the drift rate of the underlying Wiener process can be viewed as a draw from normal distribution with mean $v$ and variance $\eta$. To be explicit, the parameter $\eta$ does not have any effect on noise \textit{within the trial} as that noise is solely controlled by the underlying Wiener process (i.e., $\sigma^2$). Rather, the parameter $\eta$ controls the \textit{inter-trial} variability of the drift rate of the DDM by allowing the drift rate to change across trials. Letting $\eta > 0$ can predict slow errors (i.e., different average response times between upper and lower responses in the absence of a response bias), a data pattern that earlier evidence accumulation models could not predict \parencite{ratcliff1978theory}. Setting $\eta = 0$ removes all variance from the distribution and thus eliminates any inter-trial variability in the drift rate (i.e., the drift rate of the underlying Wiener process will always be precisely $v$). It can be verified that setting $\eta = 0$ in either of the four-parameter variable drift rate variants of the density functions, Equations \eqref{eq:den-var-s} and \eqref{eq:den-var-l}, will simplify to the three-parameter constant drift rate variants of the density functions, Equations \eqref{eq:den-con-s} and \eqref{eq:den-con-l}, respectively.

\subsubsection{Additional Notes} \label{sec:den-our-adl}

Furthermore, we exclude $t_0$ from our equations because it only serves as an additive shift in the data $t$. That is, we can remap $t := t - t_0$, and we will use this remapping in place of $t$ for the remainder of this paper.

Finally, our formulation of the density function uses the parameter $w$ as the \textit{relative} starting point of the diffusion process. Other formulations may use the parameter $z \in (0, a)$ as the \textit{absolute} starting point of the diffusion process. As $w$ is a scaled version of $z$, the relation between the two parameters is $w = \frac{z}{a}$.

\subsection{The Density Function Equations} \label{sec:den-equ}

Following the exposition in the previous subsection, we will refer to our interpretation of the DDM density function as either $f(t ~|~ v, \eta, a, w)$ or $f(t ~|~ v, a, w)$ depending on whether we are including inter-trial variability in the drift rate or not, respectively. Using the previous derivations from the literature \parencite{feller1968introduction}, there are three different versions of the probability density function: ``large-time'' with constant drift rate, $f_\ell (t | v, a, w)$, Equation \eqref{eq:den-con-l}; ``small-time'' with constant drift rate, $f_s (t | v, a, w)$, Equation \eqref{eq:den-con-s}; and ``small-time'' with variable drift rate across trials, $f_s (t | v, \eta, a, w)$, Equation \eqref{eq:den-var-s}. We will introduce another variant of the ``large-time'' density function that allows for a variable drift rate across trials, $f_\ell (t | v, \eta, a, w)$, Equation \eqref{eq:den-var-l}.

These density functions are named with a timescale label (``small'' vs. ``large'') because of their supposed advantageous computational performance for response times on the timescale of their respective labels. The categorization of a ``small'' and ``large'' response time depends on the other parameters, but in practice the threshold between these two labels is on the scale of 500ms. In this section, we will present the aforementioned variants of the density functions; each of the density function approximation methods covered later in this paper is related to one of these four variants.

\subsubsection{Constant Drift Rate Density Functions} \label{sec:den-equ-con}

We begin by considering the variant with three parameters, $v$,  $a$, and $w$ (i.e., $\eta = 0$). This variant is known as the \textit{constant} drift rate density function because the drift rate is held constant across trials due to the lack of inter-trial variability. There are two forms of the constant drift rate DDM variant: the ``large-time'' and ``small-time'' density functions. Both forms are found in the literature, and we present our formulations of them here. The ``large-time'' constant drift rate density function is
\begin{equation} \label{eq:den-con-l}
	f_\ell(t ~|~ v, a, w) = \frac{\pi}{a^2}
	e^{ \left( -vaw-\frac{v^2 t}{2} \right)}
	\sum_{j = 1}^{\infty} j \sin \left( j w \pi \right) e^{ \left( -\frac{j^2 \pi^2 t}{2a^2} \right)}.
\end{equation}
The ``small-time'' constant drift rate density function is
\begin{equation} \label{eq:den-con-s}
	f_s(t ~|~ v, a, w) = \frac{a}{\sqrt{2 \pi t^3}}
	e^{ \left( -vaw-\frac{v^2 t}{2} \right)}
	\sum_{j = -\infty}^{\infty} (w + 2j) e^{ \left( -\frac{a^2}{2t} \left( w + 2j \right)^2 \right)}.
\end{equation}

\subsubsection{Variable Drift Rate Density Functions} \label{sec:den-equ-var}

Next we expand the three-parameter variant to the four-parameter variant in order to allow the drift rate to vary across trials (i.e., $\eta > 0$). This variant is known as the \textit{variable} drift rate density function because the drift rate is allowed to vary across trials. There is currently only the ``small-time'' variable drift rate density function available in the literature, so we will present our formulation of that. In addition, we will introduce its ``large-time'' counterpart by using the equality of the two constant drift rate density functions. The ``small-time'' variable drift rate density function is
\begin{equation} \label{eq:den-var-s}
	f_s(t ~|~ v, \eta, a, w) = \frac{a}{\sqrt{2 \pi t^3 \left( 1 + \eta^2 t \right)}}
	e^{ \left( \frac{\eta^2 a^2 w^2 -2vaw -v^2 t}{2 (1 + \eta^2 t)} \right)}
	\sum_{j = -\infty}^{\infty} (w + 2j) e^{ \left( -\frac{a^2}{2t} \left( w + 2j \right)^2 \right)}.
\end{equation}

Notice that both the terms and indices in the infinite sums are identical between the two ``small-time'' models, Equations \eqref{eq:den-con-s} and \eqref{eq:den-var-s}, with only the multiplicative scalar differing. Exploiting these identical summations and using the equality of the two constant drift rate density functions Equations \eqref{eq:den-con-l} and \eqref{eq:den-con-s}, we can write a ``large-time'' version of the density function that includes inter-trial variability in the drift rate (i.e., $\eta > 0$). Detailed steps to arrive at this density function are presented in Appendix \ref{app:mat-lar}. The ``large-time'' variable drift rate density function is
\begin{equation} \label{eq:den-var-l}
	f_\ell(t ~|~ v, \eta, a, w) = \frac{\pi}{a^2 \sqrt{1 + \eta^2 t}}
	e^{ \left( \frac{\eta^2 a^2 w^2 -2vaw -v^2 t}{2 (1 + \eta^2 t)} \right)}
	\sum_{j = 1}^{\infty} j \sin \left( j w \pi \right) e^{ \left( -\frac{j^2 \pi^2 t}{2a^2} \right)}.
\end{equation}

\subsubsection{Converting Between the Constant and Variable Drift Rate Variants} \label{sec:den-equ-m}

Since the summation for each time scale (``small-time'' and ``large-time'') is the same regardless of including variability in the drift rate, it follows that there exists a term $M$ such that the density function for the constant drift rate variant multiplied by $M$ yields the density function for the variable drift rate variant. That is, $M \cdot f(t | v, a, w) = f(t | v, \eta, a, w)$ from the above equations. Note that $M$ works for converting both the ``large-time'' and ``small-time'' constant drift rate densities to variable drift rate densities. Although we do not use this term in our implementations of the approximation methods, it may be useful in adapting preexisting approximation methods to output the density with variable drift rate. It is pertinent to note, however, that this conversion may produce inaccurate numerical approximations in certain cases. For more details and an example, see the \textit{Validity Vignette} on the \texttt{fddm} \texttt{CRAN} page\footnote{\url{https://cran.r-project.org/package=fddm/vignettes/validity.html\#den-ke}}. The multiplicative term $M$ is given below:
\begin{equation} \label{eq:M}
	M = \frac{1}{\sqrt{1 + \eta^2 t}} \exp \left( vaw + \frac{v^2 t}{2} + \frac{\eta^2 a^2 w^2 -2vaw -v^2 t}{2 (1 + \eta^2 t)} \right).
\end{equation}

\graphicspath{{./Images/sec_3_approximations/}}
\section{Approximations to the DDM Density Functions} \label{sec:apx}

In order to approximate either the ``large-time'' or the ``small-time'' density functions, we must choose how to truncate the terms in the infinite sum so that the overall error in the approximation is less than the desired level of accuracy, $\epsilon$. To guarantee this precision, the existing approximation methods first calculate $k$, the number of terms required in the truncated sum so that the overall error of the approximation is less than $\epsilon$. Fortunately, the infinite sums for each time scale are identical whether or not inter-trial variability is included; thus any truncation method is equally applicable to the constant drift rate and variable drift rate density functions, scaling the desired error tolerance appropriately.

The literature provides three methods for calculating $k$. \textcite{navarro2009fast} introduce the first two methods, one for the ``large-time'' approximation and one for the ``small-time'' approximation. To maximize the efficiency of both approximations, \textcite{navarro2009fast} suggest combining the ``large-time'' and ``small-time'' approximations by minimizing the overall number of terms required in one of the summations. In this ``combined-time'' framework, the two approximations are compared by their precalculation of $k$, and the approximation that requires fewer terms is selected. The third method for calculating $k$ is provided by \textcite{gondan2014even} and is another method to determine the number of terms required in the ``small-time'' truncated sum. This newer method can also be used in the ``combined-time'' framework of \textcite{navarro2009fast} by directly substituting the original method of precalculating $k$ with the new method.

In contrast to the existing approximation methods, we can take an alternative approach by simply adding terms to the truncated ``small-time'' summation until the error of the approximation is guaranteed to be less than the given error tolerance. This way we avoid the precalculation of $k$, decreasing the number of computations that the approximation incurs. This approximation method was suggested by \textcite{gondan2014even} and is only valid in the ``small-time'' density function because the terms in this summation can be treated as an alternating and absolutely decreasing convergent sequence. This fact and a formal proof of this approximation method's validity are provided in Appendix \ref{app:mat-inf-sma}. For ease of notation, we will refer to this approximation method as \textit{SWSE} for the duration of this paper (an acronym for Stop When Small Enough). Moreover, we can implement a simple heuristic approach to use this approximation method in the ``combined-time'' framework of \textcite{navarro2009fast}, despite not precalculating $k$.

To make matters even more complicated, there exist two mathematically equivalent formulations for the ``small-time'' summation. We can pair each of these with a different method for calculating the number of terms in the truncation, yielding double the number of available ``small-time'' approximation methods. The mathematical equivalence of these two sequences is shown in Appendix \ref{app:mat-sum}, and so we will refrain from adding more detail here. The two different styles of summation are shown below in Equation \eqref{eq:sum}, named according to the year they were published by \textcite{gondan2014even} and \textcite{blurton2017first}.

\begin{equation} \label{eq:sum}
	\begin{aligned}
		S_{14} &= \sum_{j = -\infty}^{\infty} (w + 2j) e^{ \left( -\frac{a^2}{2t} \left( w + 2j \right)^2 \right)},\\[2ex]
		S_{17} &= \sum_{j = 0}^{\infty} \left( -1 \right)^j r_j e^{ \left( -\frac{a^2}{2t} r_j^2 \right)},\\
		r_j &= \begin{cases}
			j + w & \text{ if $j$ is even,}\\
			j + 1 - w & \text{ if $j$ is odd.}
		\end{cases}\\
	\end{aligned}
\end{equation}

Ultimately, there is one approximation method to the ``large-time'' density function, and there are three methods to approximate the ``small-time'' density function. However, by using both styles of summation we double the number of ``small-time'' approximation methods to six. Then combining those six ``small-time'' approximation methods with the single ``large-time'' approximation method yields six ``combined-time'' approximation methods. In total, there are thirteen approximation methods for the DDM density functions, and we will discuss each of them in the following subsections.

\subsection{Large Time} \label{sec:apx-lar}

We will begin our discussion of the many density function approximations by temporarily restricting ourselves to the ``large-time'' density function, Equation \eqref{eq:den-con-l}. The approximation of the ``large-time'' density function is ultimately a truncation of the infinite summation contained within it, as this infinite sum is technically incalculable because the analytic solution is still unknown. Since it can be shown that this infinite sum converges (see Appendix \ref{app:mat-inf-lar}), the most straightforward way to truncate this infinite sum is to only include the first $k$ terms\footnote{Note that the term with index $0$ evaluates to $0$, so we exclude this term from the summation notation to mitigate combinatorial confusion.} (i.e., the indices $\sett{1, \dots, k}$). Substituting $k$ for $\infty$ in Equations \eqref{eq:den-con-l} and \eqref{eq:den-var-l} yields the truncated version of the ``large-time'' density functions:

\begin{equation} \label{eq:apx-lar}
	\begin{aligned}
		f_\ell^{\angbrk{k}}(t ~|~ v, a, w) &= \frac{\pi}{a^2}
		e^{ \left( -vaw-\frac{v^2 t}{2} \right)}
		\sum_{j = 1}^{k} j \sin \left( j w \pi \right) e^{ \left( -\frac{j^2 \pi^2 t}{2a^2} \right)},\\
		f_\ell^{\angbrk{k}}(t ~|~ v, \eta, a, w) &= \frac{\pi}{a^2 \sqrt{1 + \eta^2 t}}
		e^{ \left( \frac{\eta^2 a^2 w^2 -2vaw -v^2 t}{2 (1 + \eta^2 t)} \right)}
		\sum_{j = 1}^{k} j \sin \left( j w \pi \right) e^{ \left( -\frac{j^2 \pi^2 t}{2a^2} \right)}.
	\end{aligned}
\end{equation}

\begin{figure}[tb!]
	\begin{center}
		\includegraphics[width=\textwidth]{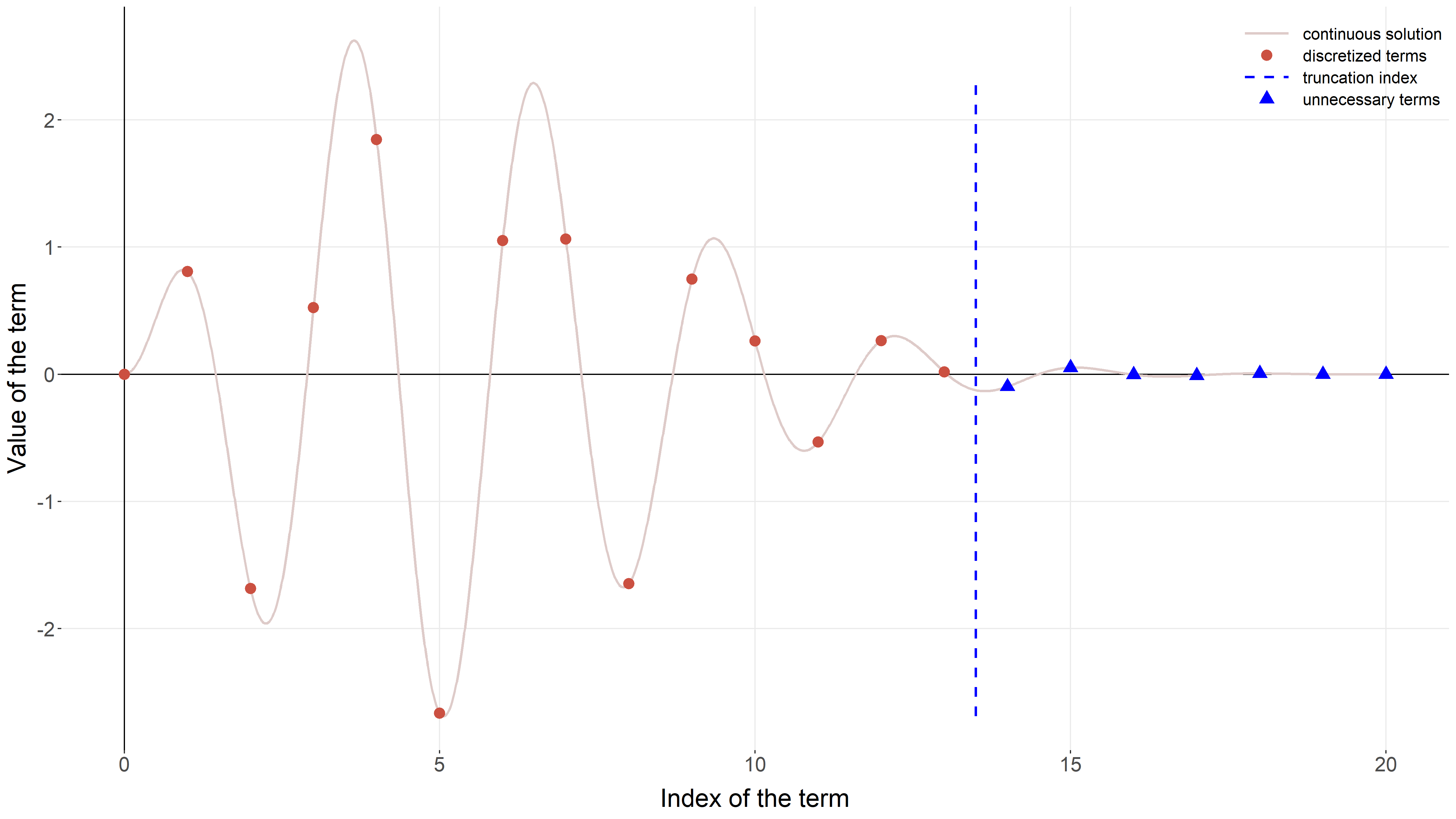}
	\end{center}
	\caption{The first twenty non-zero terms in the ``large-time'' summation, as described in Section \ref{sec:apx-lar}. On the horizontal axis is the index of the term in the summation, and the vertical axis shows the value of each individual term in the summation. As the terms of a summation are inherently discrete, the discretized terms are plotted over the underlying continuous values of the summation terms. The overall value of the truncated ``large-time'' summation is given by the sum of the values of the terms (red points) before the truncation index (vertical dashed blue line); this truncation index is purely illustrative. The sum of all values of terms whose index is greater than this truncation index (blue triangles) should be less than the desired precision, $\epsilon$.}
	\label{fig:apx-large-terms}
\end{figure}

Figure \ref{fig:apx-large-terms} shows an example of the terms in the ``large-time'' truncated sum, which display a decaying behavior. This decaying nature of the summation terms makes the summation itself converge and, thus, allows it to be approximated with sufficient accuracy.

\subsubsection{\texorpdfstring{\textcite{navarro2009fast}}{Navarro and Fuss (2009)}} \label{sec:apx-lar-nav}

There is only one available method for calculating $k_\ell$, the number of terms required in the truncated ``large-time'' summation. \textcite{navarro2009fast} provide this method, and given the desired precision, $\epsilon$, we define:
\begin{equation} \label{eq:kl_nav}
	k_\ell^\text{Nav} = \ceil{ \max \sett{\sqrt{\frac{-2 \log(\pi t \epsilon)}{\pi^2 t}}, \frac{1}{\pi \sqrt{t}}}},
\end{equation}
where $\ceil{\cdot}$ is the ceiling function; that is, the argument is rounded up to the nearest integer that is greater than or equal to the argument. Using this calculation, it is straightforward to evaluate the rest of the density functions $f_\ell^{\angbrk{k}}$ from Equation \eqref{eq:apx-lar}. We will label this approximation method as $f_\ell^\text{Nav}$.

\subsection{Small Time} \label{sec:apx-sma}

The approximation of the ``small-time'' density function is also a truncation of the infinite summation contained within it, as this infinite sum is also technically incalculable because the analytic solution is still unknown. Since it can be shown that this infinite sum converges (see Appendix \ref{app:mat-inf-sma}), the most straightforward way to truncate this infinite sum is to include only the first $k$ terms. However, as there are two summation styles for the ``small-time'' density functions, we must define how to count the first $k$ terms for each summation style.

The $S_{14}$ style summation extends to infinity in both the positive and negative directions, so we consider the first $\rfrac{k}{2}$ terms in both directions. More precisely, we include the terms whose indices are no greater than $\rfrac{k}{2}$ in absolute value (i.e., the indices $\sett{-\floor{\rfrac{k}{2}}, \dots, 0, \dots, \floor{\rfrac{k}{2}}}$). Note that $\floor{\cdot}$ indicates the floor function; that is, the argument is rounded down to the nearest integer that is less than or equal to the argument. In the formulation provided by \textcite{gondan2014even}, $k$ can only be odd so that the total number of terms is always exactly $k$. In the case of \textcite{navarro2009fast}, $k$ can also be even, resulting in $k + 1$ terms for the $S_{14}$ style truncated summation. Substituting $\floor{\rfrac{k}{2}}$ for $\infty$ in Equations \eqref{eq:den-con-s} and \eqref{eq:den-var-s} yields the truncated version of the ``small-time'' density functions with the $S_{14}$ summation style:

\begin{equation} \label{eq:apx-sma-14}
	\begin{aligned}
		f_s^{\angbrk{k}}(t ~|~ v, a, w) &= \frac{a}{\sqrt{2 \pi t^3}}
		e^{ \left( -vaw-\frac{v^2 t}{2} \right)}
		\sum_{j = -\floor{\rfrac{k}{2}}}^{\floor{\rfrac{k}{2}}} (w + 2j) e^{ \left( -\frac{a^2}{2t} \left( w + 2j \right)^2 \right)},\\
		f_s^{\angbrk{k}}(t ~|~ v, \eta, a, w) &= \frac{a}{\sqrt{2 \pi t^3 \left( 1 + \eta^2 t \right)}}
		e^{ \left( \frac{\eta^2 a^2 w^2 -2vaw -v^2 t}{2 (1 + \eta^2 t)} \right)}
		\sum_{j = -\floor{\rfrac{k}{2}}}^{\floor{\rfrac{k}{2}}} (w + 2j) e^{ \left( -\frac{a^2}{2t} \left( w + 2j \right)^2 \right)}.
	\end{aligned}
\end{equation}

In contrast, the $S_{17}$ style ``small-time'' summation only extends to infinity in the positive direction. Thus, it is straightforward to count the first $k$ terms in the summation (i.e., the indices $\sett{0, \dots, k-1}$). First, we use the $S_{17}$ summation style from Equation \eqref{eq:sum} in place of the $S_{14}$ summation style that is currently used in Equations \eqref{eq:den-con-s} and \eqref{eq:den-var-s}. Then we can substitute $k-1$ for $\infty$ in these equations to yield the truncated version of the ``small-time'' density functions with the $S_{17}$ summation style:

\begin{equation} \label{eq:apx-sma-17}
	\begin{aligned}
		f_s^{\angbrk{k}}(t ~|~ v, a, w) &= \frac{a}{\sqrt{2 \pi t^3}}
		e^{ \left( -vaw-\frac{v^2 t}{2} \right)}
		\sum_{j = 0}^{k-1} \left( -1 \right)^j r_j e^{ \left( -\frac{a^2}{2t} r_j^2 \right)},\\
		f_s^{\angbrk{k}}(t ~|~ v, \eta, a, w) &= \frac{a}{\sqrt{2 \pi t^3 \left( 1 + \eta^2 t \right)}}
		e^{ \left( \frac{\eta^2 a^2 w^2 -2vaw -v^2 t}{2 (1 + \eta^2 t)} \right)}
		\sum_{j = 0}^{k-1} \left( -1 \right)^j r_j e^{ \left( -\frac{a^2}{2t} r_j^2 \right)}.
	\end{aligned}
\end{equation}

\begin{figure}[tb!]
	\begin{center}
		\includegraphics[width=\textwidth]{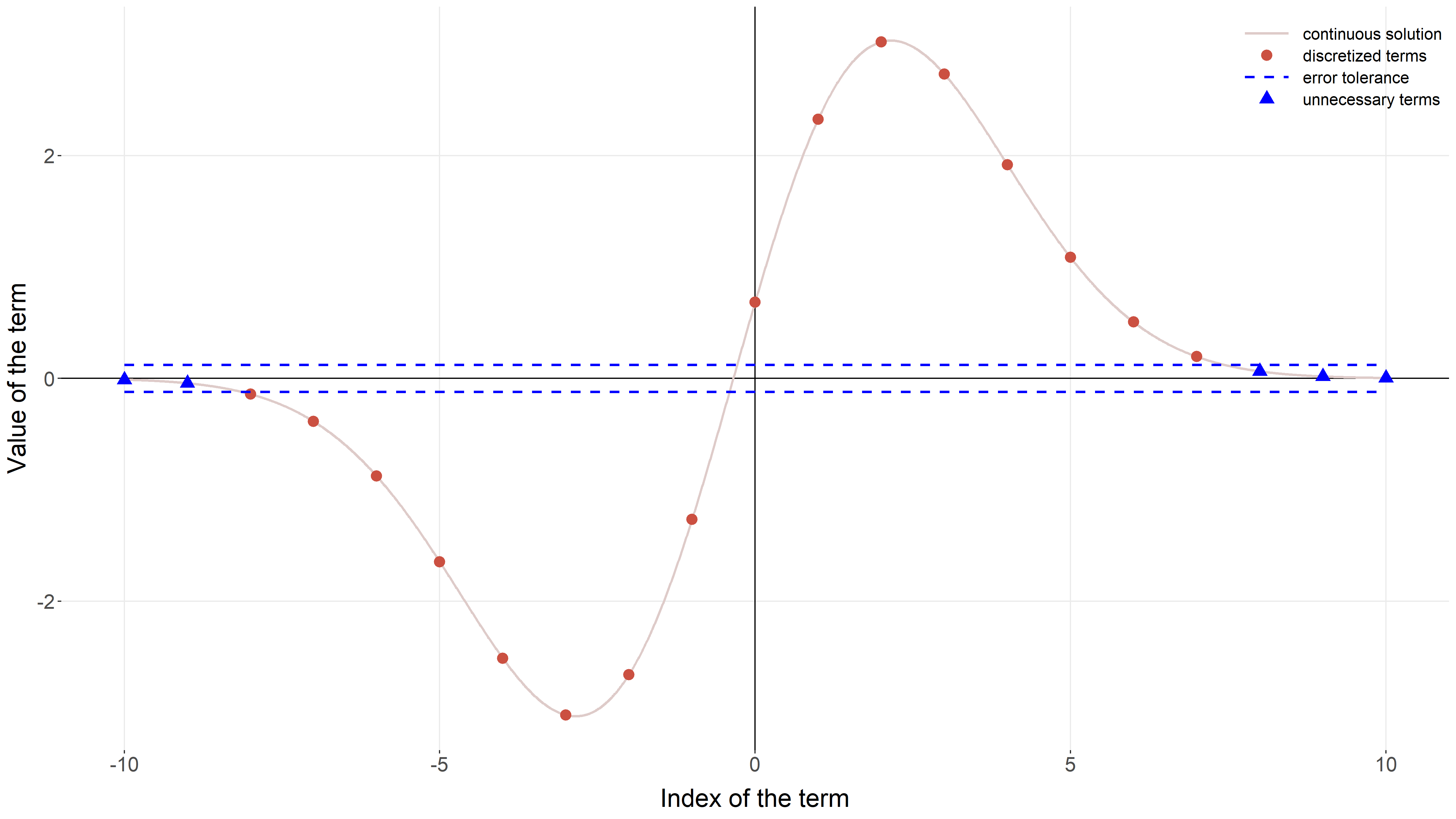}
	\end{center}
	\caption{The first twenty-one terms in the ``small-time'' summation in the $S_{14}$ style, as described in Section \ref{sec:apx-sma}. On the horizontal axis is the index of the term in the summation, and the vertical axis shows the value of each individual term in the summation. As the terms of a summation are inherently discrete, the discretized terms are plotted over the underlying continuous values of the summation terms. The overall value of the truncated ``small-time'' summation is given by the sum of the values of the terms (red points) outside of the error tolerance (horizontal dashed blue lines); this error tolerance is purely illustrative. The sum of all values of terms that lie between these two horizontal dashed lines (blue triangles) should be less than the desired precision, $\epsilon$.}
	\label{fig:apx-small-terms-14}
\end{figure}

\begin{figure}[tb!]
	\begin{center}
		\includegraphics[width=\textwidth]{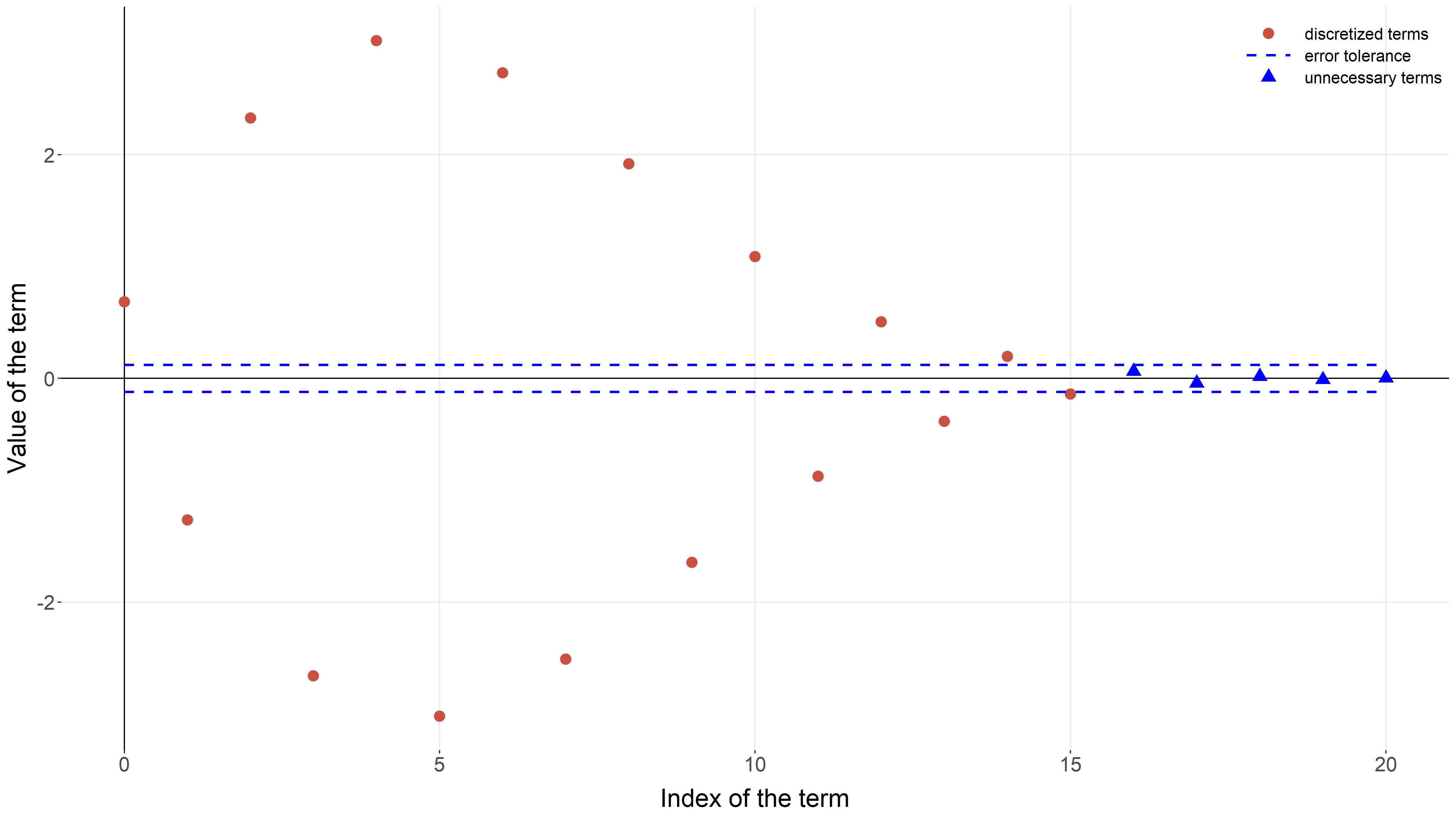}
	\end{center}
	\caption{The first twenty-one terms in the ``small-time'' summation in the $S_{14}$ style, as described in Section \ref{sec:apx-sma}. On the horizontal axis is the index of the term in the summation, and the vertical axis shows the value of each individual term in the summation. The discretized terms of the summation are plotted according to the rearrangement in Section \ref{sec:apx-sma}, and as such there is no continuous solution similar to Figure \ref{fig:apx-small-terms-14}. The overall value of the truncated ``small-time'' summation is given by the sum of the values of the terms (red points) outside of the error tolerance (horizontal dashed blue lines); this error tolerance is purely illustrative. The sum of all values of terms that lie between these two horizontal dashed lines (blue triangles) should be less than the desired precision, $\epsilon$.}
	\label{fig:apx-small-terms-17}
\end{figure}

Figure \ref{fig:apx-small-terms-14} shows an example of the terms in the ``small-time'' truncated sum using the $S_{14}$ summation style. Here, the terms appear to only show a general decay as the index gets further from zero. This decaying behavior is sufficient to demonstrate that the summation itself converges and, thus, we can approximate it with the given accuracy. However, rearranging the terms into the $S_{17}$ style (i.e., order the indices as$\sett{0, -1, 1, \dots, -i, i, -(i+1), i+1, \dots}$) as shown in Figure \ref{fig:apx-small-terms-17} reveals a distinct oscillatory behavior that allows for an even simpler approximation while maintaining the desired precision. This simpler approximation method is described in Section \ref{sec:apx-sma-swse}.

In addition to the two equivalent styles of summation from Equation \eqref{eq:sum}, there are three different methods to truncate the sum. Because of the multiple methods we will separate them into three further subsections, one per truncation method, that will each contain the two equivalent summation styles. Ultimately, we have six distinct approximation methods for the ``small-time'' density function: $f_s^\text{Nav,14}$, $f_s^\text{Nav,17}$, $f_s^\text{BGK,14}$, $f_s^\text{BGK,17}$, $f_s^\text{SWSE,14}$, and $f_s^\text{SWSE,17}$.

\subsubsection{\texorpdfstring{\textcite{navarro2009fast}}{Navarro and Fuss (2009)}} \label{sec:apx-sma-nav}

First, \textcite{navarro2009fast} provided a method for calculating the required number of terms in the truncated ``small-time'' summation. Given a desired precision $\epsilon$, they define
\begin{equation} \label{eq:ks_nav}
	k_s^{\text{Nav}} = \ceil{ \max \sett{2 + \sqrt{-2t \log(2 \epsilon \sqrt{2 \pi t})}, 1 + \sqrt{t}}},
\end{equation}
where $\ceil{\cdot}$ is the ceiling function, as mentioned previously. From this calculation, the evaluation of the density function approximations can proceed with either summation style $S_{14}$ or $S_{17}$, Equations \eqref{eq:apx-sma-14} and \eqref{eq:apx-sma-17}.

\subsubsection{\texorpdfstring{\textcite{gondan2014even}}{Gondan et al. (2014)}} \label{sec:apx-sma-gon}

After \textcite{navarro2009fast} published their paper, \textcite{gondan2014even} introduced another method for calculating the required number of terms in the truncated ``small-time'' summation. It is important to note, however, that \textcite{gondan2014even} provided the number of required \textit{pairs} of terms in the $S_{14}$ summation style, and not the number of required \textit{individual} terms. As we want the number of \textit{individual} terms, we adapt their formula in Equation \eqref{eq:ks_gon}. Given a desired precision $\epsilon$, we define
\begin{equation} \label{eq:ks_gon}
	\begin{aligned}
		k_s^{\text{Gon}} &= 1 + 2 \cdot \ceil{ \max \sett{ \tfrac{1}{2} \left( \sqrt{2t} - w \right), \tfrac{1}{2} \left( \sqrt{-t (u_\epsilon - \sqrt{-2 u_\epsilon -2})} - w \right)}},\\
		u_\epsilon &= \min \sett{-1, \log(2 \pi t^2 \epsilon^2)},
	\end{aligned}
\end{equation}
where $\ceil{\cdot}$ is again the ceiling function. Similarly to the previous method, the density function approximations can proceed with either summation style $S_{14}$ or $S_{17}$, Equations \eqref{eq:apx-sma-14} and \eqref{eq:apx-sma-17}.

\subsubsection{SWSE (Stop When Small Enough)} \label{sec:apx-sma-swse}

Since both of the ``small-time'' summation styles can be treated as convergent absolutely decreasing series, we can use our SWSE method of truncating the sum. This approximation method was mentioned by \textcite{gondan2014even}, but they opted to only formally present their novel calculation of $k$ for the ``small-time'' truncated summation. The general idea of the SWSE approximation method is to take advantage of the alternating and decreasing nature of the terms in the infinite sum to bound the approximation's error. After arranging the terms as shown by \textcite{gondan2014even}, we can apply the alternating series test to place an upper bound on the truncation error after including a certain number of terms in the summation. This upper bound is the absolute value of the next term in the sequence; thus we can truncate the infinite sum once one of its terms is absolutely less than the desired precision, $\epsilon$. The validity of this approximation method is proven in Appendix \ref{app:mat-inf-sma}. As with the two other ``small-time'' approximation methods, either style of summation ($S_{14}$ or $S_{17}$) can be used which results in two more approximation methods for the ``small-time'' density functions.

\subsection{Combining Small and Large Time} \label{sec:apx-mix}

\textcite{navarro2009fast} suggest a strategy that uses a combination of the ``large-time'' and ``small-time'' approximations, based on whichever approximation is more efficient. Unfortunately, the mechanism that they provide is a quasi-nondeterministic function so we slightly change its practical implementation, as described in Section \ref{sec:imp-k}. Nonetheless, the concept is logical, and we provide six possible ``combined-time'' approximation methods based on their mechanism as there are six ``small-time'' approximation methods and one ``large-time'' approximation method. These approximation methods largely depend on the precalculation and subsequent comparison of $k_\ell$ and $k_s$, the required number of terms in the truncated versions of the ``large-time'' and ``small-time'' summations, respectively. After precalculation, either the ``large-time'' or ``small-time`` approximation is used, based on whichever approximation requires fewer terms. The exceptions to this original mechanism are the two ``combined-time'' approximation methods that leverage the SWSE ``small-time'' approximation method, as these only require the precalculation of $k_\ell$ and not $k_s$.

\subsubsection{\texorpdfstring{\textcite{navarro2009fast}}{Navarro and Fuss (2009)}} \label{sec:apx-mix-nav}

First, we can use the method for calculating $k_s$ as provided by \textcite{navarro2009fast} as the representation of efficiency for the ``small-time'' approximation. By precalculating both $k_\ell^{\text{Nav}}$ and $k_s^{\text{Nav}}$, we can then determine which timescale is more efficient. Given a desired precision $\epsilon$, they define
\begin{equation} \label{eq:km_nav}
	k_c^{\text{Nav}} = \min \sett{k_\ell^{\text{Nav}}, k_s^{\text{Nav}}}.
\end{equation}
From this calculation, the evaluation of the density function approximation can proceed with either summation style $S_{14}$ or $S_{17}$, Equations \eqref{eq:apx-sma-14} and \eqref{eq:apx-sma-17}, yielding two ``combined-time'' approximation methods.

\subsubsection{\texorpdfstring{\textcite{gondan2014even}}{Gondan et al. (2014)}} \label{sec:apx-mix-gon}

Although this approximation method has not been explored in the literature, we can implement the more recent method of calculating $k_s$ from \textcite{gondan2014even} in the same manner to that of \textcite{navarro2009fast}. Precalculating and comparing $k_\ell^\text{Nav}$ and $k_s^\text{Gon}$ allows us to determine which timescale is more efficient. Given a desired precision $\epsilon$, we define
\begin{equation} \label{eq:km_gon}
	k_c^{\text{Gon}} = \min \sett{k_\ell^{\text{Nav}}, ~ k_s^{\text{Gon}}}.
\end{equation}
Similarly to the previous approximation methods, the density function approximation can proceed with either summation style $S_{14}$ or $S_{17}$, Equations \eqref{eq:apx-sma-14} and \eqref{eq:apx-sma-17}, yielding two more ``combined-time'' approximation methods.

\subsubsection{SWSE (Stop When Small Enough)} \label{sec:apx-mix-swse}

The previous ``combined-time'' approximation methods have both relied on precalculating the required number of terms in both the ``large-time'' and ``small-time'' versions of the truncated sum, $k_\ell$ and $k_s$, respectively. This raises an issue with our SWSE approximation method because it does not precalculate the required number of terms in the ``small-time'' truncated sum, $k_s$. However, we can still use it with the ``large-time'' approximation method of \textcite{navarro2009fast} by implementing a heuristic switching mechanism. This mechanism will still precalculate $k_\ell^{\text{Nav}}$, but rather than comparing $k_\ell^{\text{Nav}}$ with a precalculation of $k_s$, it instead compares $k_\ell^{\text{Nav}}$ to a user-defined value, $\delta$.\footnote{argument \texttt{max\_terms\_large} in function \texttt{dfddm()}} If $k_\ell^{\text{Nav}} \leq \delta$, then the \textcite{navarro2009fast} ``large-time'' approximation method will be used; if $k_\ell^{\text{Nav}} > \delta$, then the SWSE ``small-time'' approximation method will be used. As with the other ``combined-time'' approximation methods, we can use both ``small-time'' summation styles ($S_{14}$ or $S_{17}$) to yield an additional two ``combined-time'' approximation methods.

\section{Implementation} \label{sec:imp}

In addition to introducing a novel approximation method for the DDM density function, we wish to determine the fastest such approximation method. While there have been several implementations of these approximation methods, they have largely been in different programming languages \parencite{voss2008fast, navarro2009fast, gondan2014even, rtdists, RWiener}. In order to investigate the relative speeds of the proposed approximation methods, we translated them into \texttt{C++} so that their efficiencies could be directly compared using the \texttt{dfddm()} function from the \texttt{fddm} software.

Each approximation method from the literature was provided with an implementation in a programming language, namely \texttt{R} \parencite{gondan2014even} and \texttt{MATLAB} \parencite{navarro2009fast}. Even though we wanted to use the implementations in \texttt{R}, we translated them to \texttt{C++} since it is known to be a comparatively faster programming language. To maintain the user interface via \texttt{R}, we accessed the implementations using the \texttt{R} package \texttt{Rcpp} \parencite{Rcpp1, Rcpp2, Rcpp3}. Since we were able to exploit the minimal overhead of function calls in \texttt{C++}, we wrote the code using a modular structure. This modular construction has three main advantages: first, it allow the user to select exactly which approximation method to use via one flexible interface, function \texttt{dfddm()}; second, it makes the code more interpretable by other researchers; and third, the code is easily adaptable to future additions.

For example, this modularity in the code makes it simple to switch between the two summation styles ($S_{14}$ and $S_{17})$, as the user only needs to change one function parameter that indicates which summation sub-function should be used in the approximation. Not only does this modularity make for simpler user commands to compare different approximation methods, but it also allows for the straightforward inclusion of future summation styles to the software.

\subsection{Obtaining \texorpdfstring{$k$}{k} for Multiple Observations} \label{sec:imp-k}

Our only major deviation from the original code implementations was the way in which we calculated $k$, the number of terms required in the truncated sum. Every implementation assumes an input of binary choices and the associated response times as a vector of tuples, so let that vector have length $N$. Denote the output of the function that determines the number of terms required for the infinite sum by $k_i$ for the $i^\text{th}$ response time in the input vector. In the existing implementations \parencite{navarro2009fast, gondan2014even}, every $k_i$ would be calculated and the greatest value would be used for every response time input, that is $k = \max \{k_1, k_2, \dots, k_N\}$. However this creates a function such that the output for a particular input can be different depending on the other inputs.

Consider a case where only one response time is input and it requires $k_1 = 5$ terms in the density function approximation. Now include an additional response time that requires $k_2 = 10$ terms in the approximation. In the original code implementation, the density function would use $k = \max \{5, 10\} = 10$ terms for both inputs in the density function approximation. Since the additional terms are necessarily nonzero (although may be very small in absolute value), this means that the approximation for the first response time will be necessarily different depending on whether or not the second response time input is included in the call to the density function. Using more terms than required in the truncated sum only makes the the approximation more accurate; however, we believe that this quasi-nondeterministic behavior is potentially detrimental to the reproducibility of results.

To avoid this undesirable behavior we simply do not take the maximum of the $k_i$ values and instead use each $k_i$ for its respective $i^\text{th}$ response time. Using this method ensures that the approximation is not affected by other extraneous inputs. In the example provided, the density function would use $k_1 = 5$ terms for the first input and $k_2 = 10$ terms for the second input. There is no increase in computational complexity by using this individual method instead of the maximum method because both methods only calculate each $k_i$ once.

\section{Benchmark Testing} \label{sec:ben}
\graphicspath{{./Images/sec_5_benchmark/}}

As one of our goals is to determine the fastest implementation to accurately approximate the density function, we performed benchmark tests to record the relative speed of each implementation. We compared these benchmark times across the thirteen implementations discussed in Section \ref{sec:apx} in addition to including three \texttt{R} functions from the literature that are commonly used for calculating the same PDF \parencite{gondan2014even, RWiener, rtdists}. To determine which implementations are the most computationally efficient, we test them in two environments. First, we collect benchmark data on the speeds of the implementations across a predefined parameter space. Second, we collect benchmark data on the speeds of the implementations when using them for fitting real-world data using the maximum likelihood method and a standard gradient-descent optimization algorithm; this test is perhaps of more practical value than the first test as the DDM is widely used in parameter estimation.

Since all implementations we are testing are in \texttt{R}, we use the \texttt{microbenchmark} function from the \texttt{microbenchmark} \texttt{R} package \parencite{microbenchmark} to measure their speeds. This function is a common way to collect data on the runtime of functions that can be called from within \texttt{R} since it not only performs warm up trials for each implementation but also mitigates the impact of computational noise (e.g., housekeeping of the operating system) on the results by repeating each function call several times in a randomized order.

Before we benchmark the implementations, we must first decide how to choose the default value for our switching mechanism. In line with how we are benchmarking the implementations themselves, we run benchmark tests on the ``combined-time'' implementation using different values of $\delta$ in the two environments discussed earlier in this section.

The remainder of this section is split into three subsections. First, we will go over the methods for running the different styles of benchmark tests. Second, we will discuss how we chose the default value of $\delta$. Third, we present the benchmark results for the implementations by themselves and also in a data fitting scenario. An extended set of benchmark analyses is provided in the Supplemental Materials document on GitHub.\footnote{The Supplemental Materials document can be found in the folder \texttt{paper\_analysis} of: \url{https://github.com/rtdists/fddm}}

\subsection{Methods} \label{sec:ben-met}

The methods that we discuss here are applicable to both of the following subsections; the major difference between the benchmark tests in the two subsections is the implementations fed into the benchmark tests. This subsection is broken down into two further subsections: the first part explains the method for benchmarking the implementations on a predefined parameter space, and the second part explains the method for benchmarking the implementations in the data fitting scenario.

\subsubsection{Predefined Parameter Space} \label{sec:ben-met-par}

The first type of benchmark testing simply evaluates a selection of implementations in a predefined parameter space in identical computing environments. The idea behind this type of benchmark testing is to measure the relative speeds of the implementations themselves as accurately as possible, without any distractors such as a data fitting routine or a different programming language. Implementations are selected not only based on what we want to benchmark but also the available options.

The key input to this first type of benchmark testing is the parameter space. We use two different parameter spaces in various parts of our benchmark testing, and these can be found fully defined in Table \ref{tbl:parspace-30} and Table \ref{tbl:parspace-2}. Table \ref{tbl:parspace-2} contains response times that range from $0$ seconds to $2$ seconds. As the DDM is meant to be used for decisions on the timescale of about $2$ seconds or less, this parameter space is likely the more practical of the two. However, we also consider an expanded range of response times in Table \ref{tbl:parspace-30}, because the DDM can be applied to situations where larger response times can occasionally occur. To ensure that we cover these use cases, the expanded set of response times in Table \ref{tbl:parspace-30} ranges from $0$ seconds to $30$ seconds.

An important distinction in the benchmarking methods is how to input the response times from the parameter space into the implementation. As each implementation will accept one or more response times, we have the choice to either input all the response times as a vector or input the response times individually. The most practical option is to input all of the response times as a vector since this is likely how a researcher would input the data and also how it would be done in a data fitting scenario. In addition, inputting all of the response times as a vector allows the implementations to handle vectorization and avoids the computationally expensive repeated call of the PDF functions through \texttt{R} (i.e., the overhead of calling a function in \texttt{R} compared to, say \texttt{C++}, is considerable). By inputting the response times individually, however, we are able to get a more granular parameter space in the output; this enables us to see how varying the response time can impact the benchmark data.

For our benchmark tests, we visit each point in the predefined parameter space and record the execution times for each implementation 10,000 times inside of the \texttt{microbenchmark} function. Of these 10,000 benchmark data, we only record the median in an effort to eliminate the influence of computational noise on the results. For expositional purposes, we will simply use the term ``benchmark times'' instead of ``median benchmark times'' when discussing the results of the benchmark tests.

The methods used in the following subsections will follow this template, but we will elucidate the selection of implementations, the selection of parameter space, and whether or not the response times from the parameter space were input as a vector.

\subsubsection{Data Fitting} \label{sec:ben-met-fit}

The second type of benchmark testing measures the speed of optimization routines that use a selection of implementations in the underlying likelihood functions. This type of benchmark testing is perhaps more practical than the first type as the DDM is commonly used for parameter estimation on a set of experimental results. Again, implementations are selected based on what we want to benchmark and the available options. For example, the implementation in \texttt{RWiener} and the implementation provided by \textcite{gondan2014even} do not include inter-trial variability in the drift rate; thus we cannot include them in benchmark testing that pertains to $\eta$.

The optimization routine that we use is the \texttt{R} function \texttt{nlminb} that uses the maximum likelihood method and a standard gradient-descent optimization algorithm. Since \texttt{nlminb} minimizes the likelihood function instead of maximizing it, we define a family of likelihood functions that return the negative sum of the log-likelihoods of the data. In particular, each likelihood function in this family uses a different implementation to evaluate the log-likelihoods of the data. This way we can use the same optimization process for each implementation, and the only difference in benchmark data arises from the underlying implementations that approximate the DDM density function.

As encountering local optima is a known problem for deterministic optimization algorithms, we run \texttt{nlminb} with eleven different sets of initial values. These sets of initial values are consistent throughout all of our analyses, regardless of which underlying density function approximation is being used in the optimization. The full set of initial values can be found in the supplemental materials, and there are two restrictions to these initial values that we must address. First, the parameter $t_0$ must be less than the response time, so we set the initial values for $t_0$ to be strictly less than the minimum response time for to each individual in the input data frame. Second, we must place a lower bound on $a$ that is necessarily greater than zero because the optimization algorithm occasionally evaluates the log-likelihood functions (and thus the underlying density function approximations) using values of $a$ equal to its bounds. In the case where optimization algorithm evaluates using $a = 0$, both the density function from the \texttt{rtdists} package and every implementation from \texttt{dfddm()} do not evaluate. In common use this is not an issue because very small values of $a$ do not make any sense with regard to the psychological interpretation of the parameter, but this issue can arise in an exploratory optimization environment.

Since \texttt{nlminb} is a deterministic optimization algorithm, we record various results from its optimization process: the number of calls to the likelihood function, the convergence code (i.e., either 0 for success or 1 for failure), and the value of the minimized likelihood function (known as the ``objective''). We also place upper and lower bounds on the fitted parameter values to be consistent with our interpretation of the DDM. For each data set, we collected benchmark data from five runs of the data fitting routine inside of the \texttt{microbenchmark} function. From these five runs, we again record only the median time in an effort to eliminate the influence of computational noise; again, we will use the term ``benchmark times'' instead of ``median benchmark times'' when discussing the results of the benchmark tests.

The data used for this fitting came from \textcite{trueblood2018impact}, and it contained response time data for 37 individuals. Here we considered only the data from their accuracy condition which provided 200 observations per participant. We used the parameter-fitting criteria described above to fit the parameters $a$, $v$, $w$, $t_0$ and $\eta$ separately for each of the 37 individuals included in the data. We estimated one $v$ for each stimulus class that maps onto each of the two response boundaries (i.e., ``upper'' and ``lower''), so that we estimated two drift rates and in total six free parameters per participant.

\subsection{Determining the Default Behavior of our Heuristic Switching Mechanism, \texorpdfstring{$\delta$}{delta}} \label{sec:ben-mtl}

In Section \ref{sec:apx-mix-swse}, we introduced the idea of a heuristic mechanism that would switch between use of the \textcite{navarro2009fast} ``large-time'' approximation and the SWSE ``small-time'' approximation. The main issue with combining these two approximations is that the SWSE approximation does not precalculate the number of terms, which is the basis of previous switching mechanisms. Instead, we use a user-defined value $\delta$ to determine which approximation will be used given the other parameter values. More precisely, we precalculate $k_\ell^{\text{Nav}}$ and then compare this value to $\delta$. If $k_\ell^{\text{Nav}} \leq \delta$, then the \textcite{navarro2009fast} ``large-time'' approximation will be used; if $k_\ell^{\text{Nav}} > \delta$, then the SWSE ``small-time'' approximation will be used.

The naturally ensuing problem is to find the optimal value of $\delta$. Although the optimum value for switching depends on the other parameters input to the model, we fix $\delta$ for all inputs because we treat it as a simple, heuristic value that works well enough for most practical cases. The remainder of this section will detail the methods and results that we used to justify our choice of default value, $\delta = 1$.

In order to find the best default value for $\delta$, we consider several candidate values. Note that it only makes sense for $\delta$ to be a non-negative integer because the only possible values for $k_\ell$ are non-negative integers. Based on $k_s$ and $k_\ell$ calculations at the \texttt{dfddm()} default error tolerance of $0.000001$, we tested the integers $\sett{0, 1, 2, 3, 4, 5, 6, 7}$ using both the $S_{14}$ and $S_{17}$ summation styles; this yields sixteen candidate implementations. Notice that $\delta = 0$ indicates that the SWSE ``small-time'' approximation will always be used.

\subsubsection{Predefined Parameter Space} \label{sec:ben-mtl-par}

First, we collected benchmark data on the speeds of the sixteen candidate ``combined-time'' SWSE implementations across a parameter space with response times ranging from $0$ to $30$ seconds. The response times were input as a vector to the implementations, and this parameter space can be found in Table \ref{tbl:parspace-30}. Whereas this range may seem quite large, we wanted to cover as many use cases as possible when choosing the default for our heuristic method.

\begin{table}[tb!]
	\centering
	\small
	\caption{Parameter space with response times ranging from $0$ to $30$ seconds, used for determining $\delta$.}
	\label{tbl:parspace-30}
	\begin{threeparttable}
		\begin{tabular}{cl}
			\toprule
			Parameter & Values\\
			\midrule
			t & 0.001, 0.1, 1, 2, 3, 4, 5, 10, 30\\
			a  & 0.25, 0.5, 1, 2.5, 5\\
			v  & -5, -2, 0, 2, 5\\
			w  & 0.2, 0.5, 0.8\\
			$\eta$ & 0, 0.5, 1, 1.5\\
			\bottomrule
		\end{tabular}
		\begin{tablenotes}[para, flushleft]
			\emph{Note.} Parameter $t_0$ is fixed to $0.0001$ for consistency with the other reported benchmarks.
		\end{tablenotes}
	\end{threeparttable}
	\normalsize
\end{table}

\begin{figure}[tb!]
	\begin{center}
		\includegraphics[width=\textwidth]{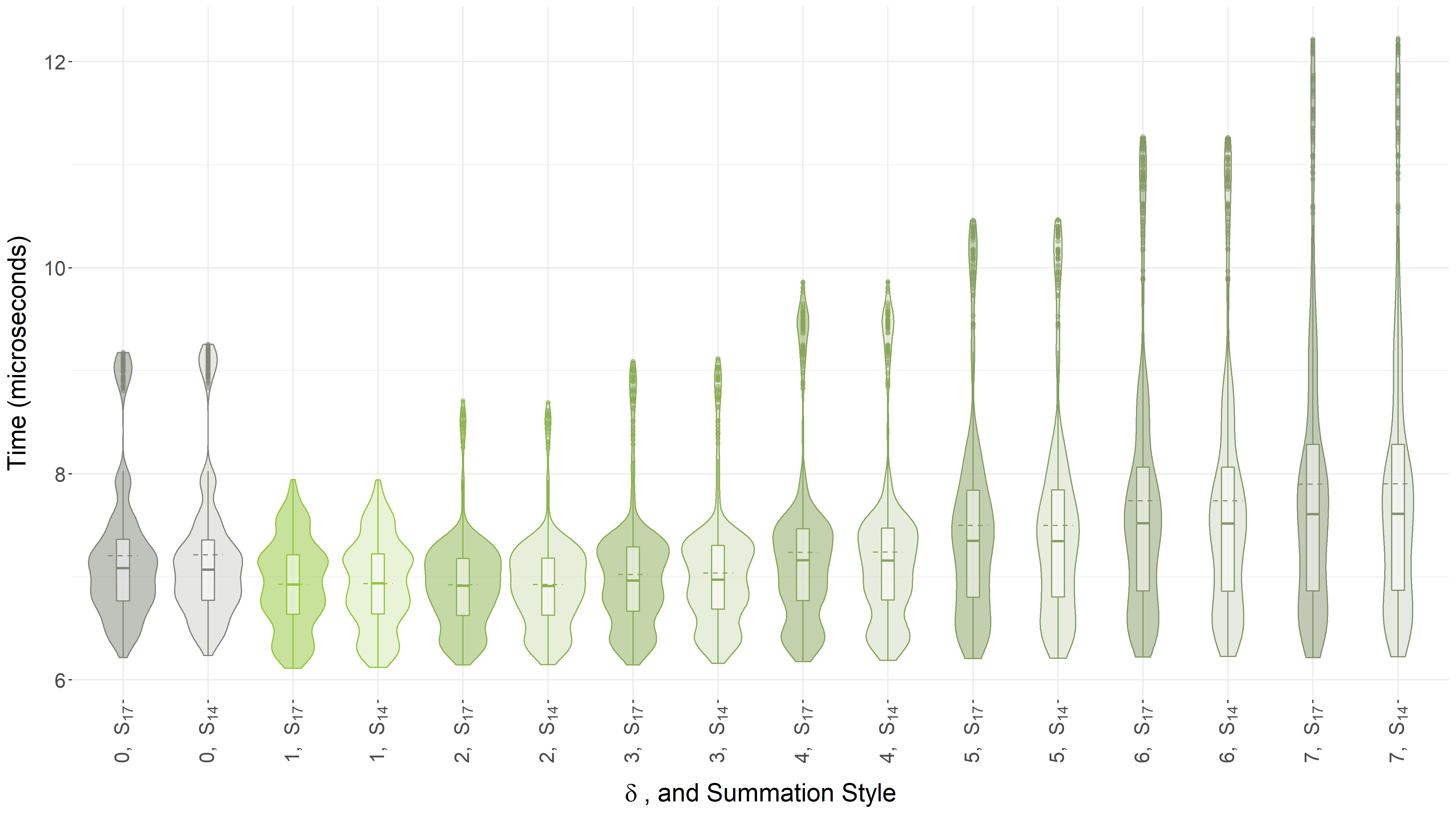}
	\end{center}
	\caption{Effect of $\delta$ on the benchmark times for the ``combined-time'' SWSE implementation and selected response times. The horizontal axis shows the combination of the chosen value of $\delta$ and the summation style used, and the vertical axis shows the benchmark time. The violin plot shows a mirrored density estimate; overlaying the violin plot, the boxplot shows the median in addition to the first and third quartiles; the horizontal dashed line shows the mean. The response times used as inputs to the implementations were between $0$ and $30$ seconds and input to the density function as a vector; the full parameter space can be found in Table \ref{tbl:parspace-30}.}
	\label{fig:ben-mtl-vec-bm}
\end{figure}

We show the results of the benchmark tests on the predefined parameter space in Figure \ref{fig:ben-mtl-vec-bm}. Immediately apparent is that there is little difference between the two summation styles, $S_{14}$ and $S_{17}$ in these benchmark data, so we will not distinguish between the two in our further analysis of the optimal value for $\delta$. We also see that $\delta = 0$ is a problematic value because it is essentially just a slower version of the SWSE ``small-time'' implementation, since it does not use the \textcite{navarro2009fast} ``large-time'' approximation yet still calculates $k_\ell^{\text{Nav}}$. Consistent with the results reported below in Section \ref{sec:ben-imp-par}, this variant has a long tail because the SWSE ``small-time'' implementation is slow for large effective response times.

It appears that relying more on the \textcite{navarro2009fast} ``large-time'' approximation (i.e., higher value of $\delta$) increases the length of the tail in the benchmark results. This is likely because the switching mechanism uses the ``large-time'' approximation when the ``small-time'' approximation is more efficient. The value $\delta = 1$ appears to be the fastest using these benchmark data. Setting $\delta = 1$ means that we almost exclusively use the SWSE ``small-time'' approximation, except when the \textcite{navarro2009fast} ``large-time'' approximation is exceptionally efficient.

\subsubsection{Data Fitting} \label{sec:ben-mtl-fit}

Second, we used the sixteen candidate ``combined-time'' SWSE implementations in a simple data fitting scenario. This analysis will not only help to determine the practical performance of each candidate value of $\delta$, but it will also highlight any potential issues with certain candidate values.

\begin{figure}[htb!]
	\begin{center}
		\includegraphics[width=\textwidth]{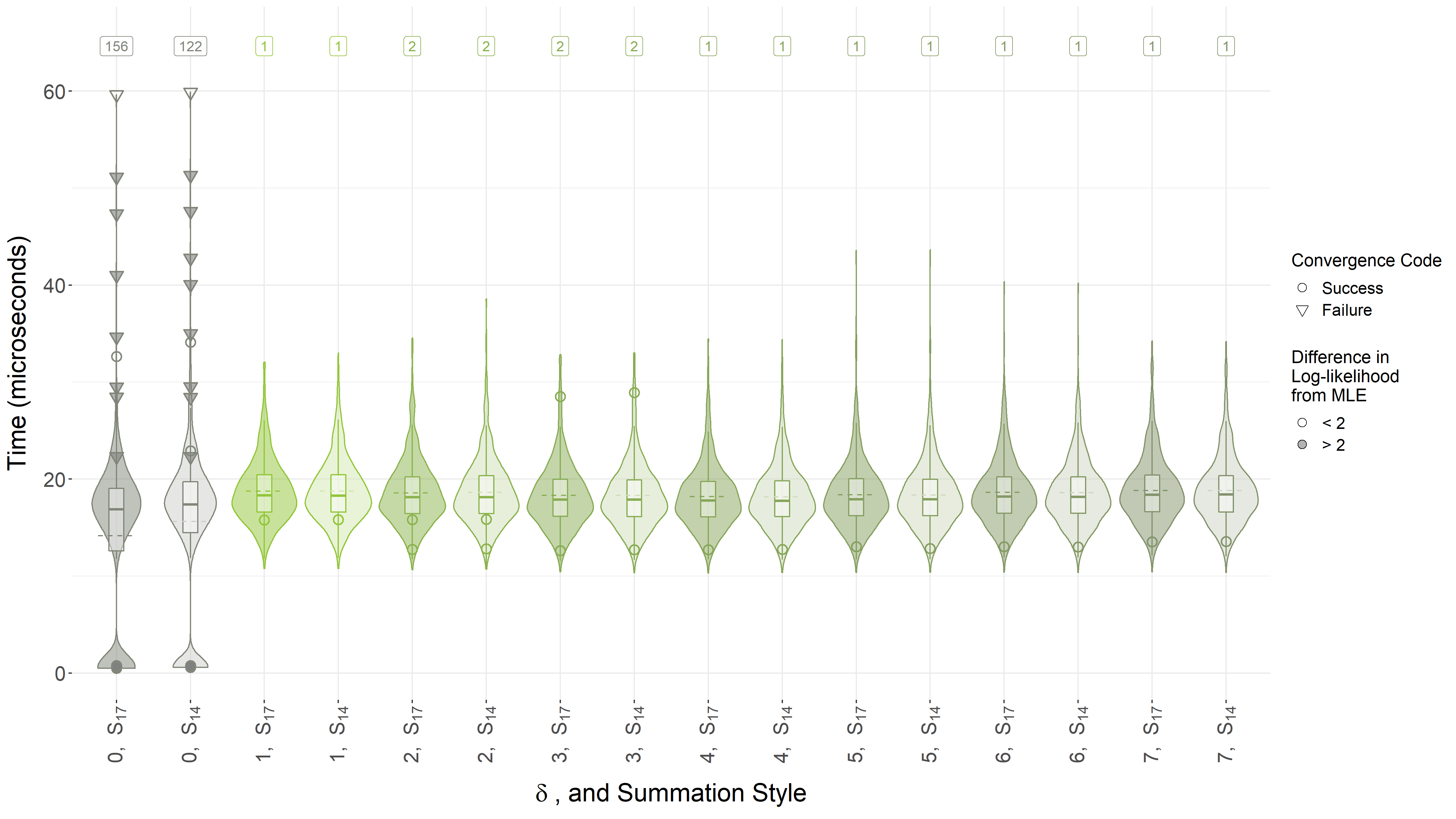}
	\end{center}
	\caption{Effect of $\delta$ on the benchmark times for the ``combined-time'' SWSE implementation during data fitting. Each distribution shows results from 407 fitting runs (the 37 participants of \nptextcite{trueblood2018impact} times 11 fixed sets of starting values). All visible points have a difference in log-likelihood greater than $0.0001$ compared to the maximum likelihood estimate for that participant; above each violin plot is the total number of such data points. ``Convergence Code'' in the legend refers to the return value provided by the optimization algorithm \texttt{nlminb}, where $0$ indicates successful convergence and $1$ indicates failed convergence. More details on the data fitting process are included in Section \ref{sec:ben-met-fit}, and more details on the graphical elements are given in Figure \ref{fig:ben-mtl-vec-bm}.}
	\label{fig:ben-mtl-fit-bm}
\end{figure}

We show the results of the benchmark tests on this data fitting scenario in Figure \ref{fig:ben-mtl-fit-bm}. Similarly to Figure \ref{fig:ben-mtl-vec-bm} above, there is little difference between the $S_{14}$ and $S_{17}$ summation styles. Moreover, we again see the shortcomings of $\delta = 0$ (i.e., just the SWSE ``small-time'' implementation) because it struggles with large response times. Specifically, we see that in many cases the optimization algorithm does not return the maximum-likelihood estimate and/or reports optimization failures. Since large response times often arise in a typical optimization routine, we should certainly pick a default value $\delta > 0$ to avoid the shortcomings of the SWSE ``small-time'' implementation. We notice that the benchmark times stay relatively consistent across the other values of $\delta$, but the tails of the distribution of benchmark times generally elongate with a higher value of $\delta$. When considering the stability of the implementations for $\delta > 0$, only $\delta \in \sett{2, 3}$ are slightly less stable than the other values. Taken together, the two benchmarks results show that $\delta = 1$ is both the fastest option as well as the most stable option. Consequently, we use $\delta = 1$ as the default value going forward.

\subsection{Benchmarking all Implementations} \label{sec:ben-imp}

Now that we have established the default behavior of our ``combined-time'' SWSE implementation, we can proceed to compare it against other implementations. In order to determine the most efficient and stable implementation, we will begin by performing benchmark tests on predefined parameter spaces. Following that, we will again use a simple data fitting scenario to test the speeds of the implementations in a practical setting.

\subsubsection{Predefined Parameter Spaces} \label{sec:ben-imp-par}

First, we collected benchmark data on the speeds of all of the implementations discussed in this paper. We measured their speeds across a practical parameter space with response times ranging from $0$ to $2$ seconds. This parameter space is perhaps more practical than the other one that has a larger range of response times because the DDM is meant to be applied to situations where the response times are two seconds or faster. The response times from this parameter space were input as a vector to the implementations, and the parameter space can be found in Table \ref{tbl:parspace-2}.

\begin{table}[tb!]
	\centering
	\small
	\caption{The parameter space used in the benchmark tests.}
	\label{tbl:parspace-2}
	\begin{threeparttable}
		\begin{tabular}{cl}
			\toprule
			Parameter & Values\\
			\midrule
			t & 0.1, 0.2, 0.3, 0.4, 0.5, 0.6, 0.7, 0.8, 0.9, 1.0, 1.1, 1.2, 1.3, 1.4, 1.5, 1.6, 1.7, 1.8, 1.9, 2.0\\
			a  & 0.5, 1, 1.5, 2, 2.5, 3, 3.5\\
			v  & -5, -2, 0, 2, 5\\
			w  & 0.3, 0.4, 0.5, 0.6, 0.7\\
			$\eta$ & 0, 1, 2, 3.5\\
			\bottomrule
		\end{tabular}
		\begin{tablenotes}[para, flushleft]
			\emph{Note.} The parameter $t_0$ is fixed to $0.0001$ because this parameter must be greater than zero for the density function from the \texttt{RWiener} package.
		\end{tablenotes}
	\end{threeparttable}
	\normalsize
\end{table}

\begin{figure}[htb!]
	\begin{center}
		\includegraphics[width=\textwidth]{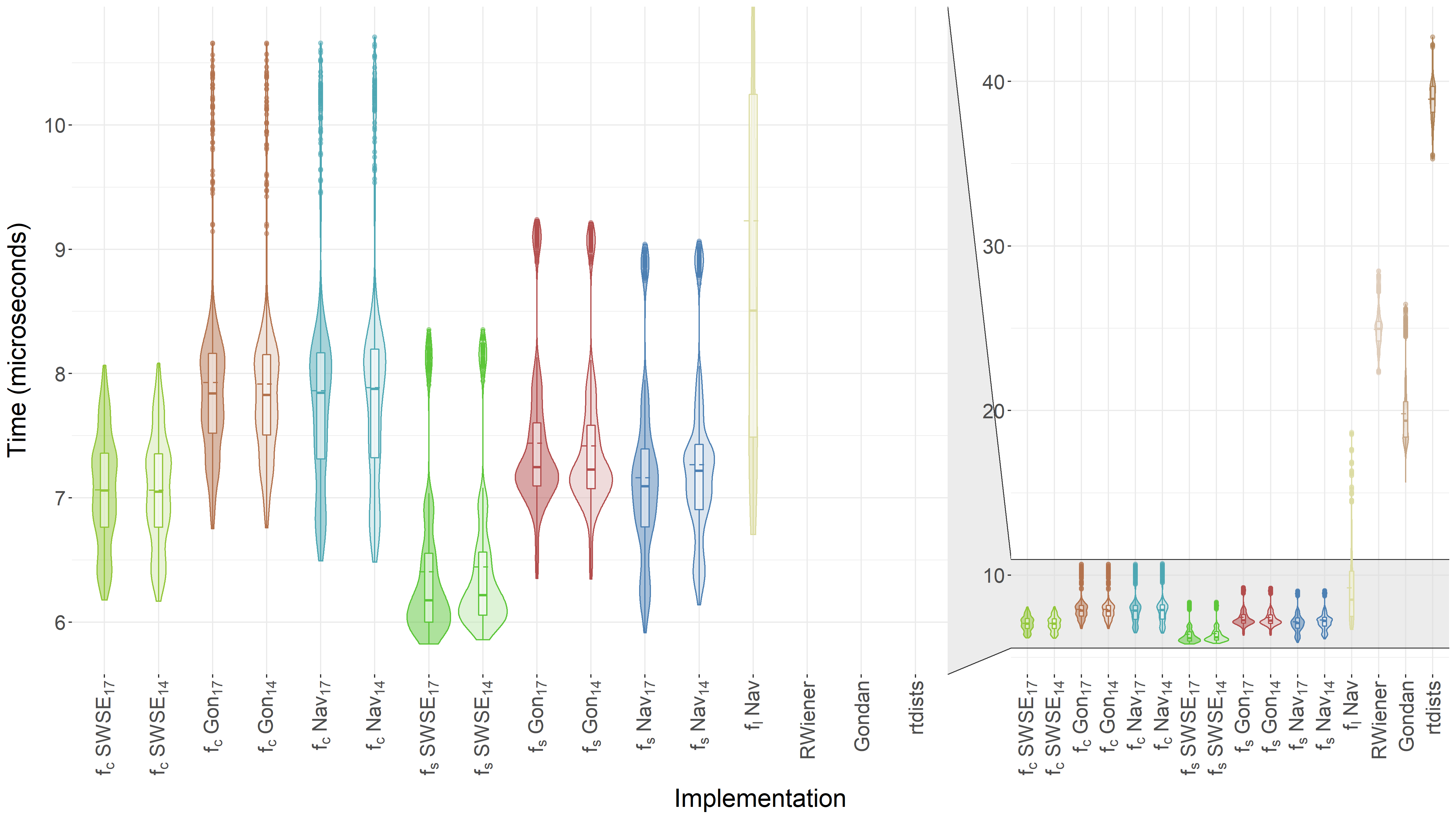}
	\end{center}
	\caption{Benchmark times of different density function approximation implementations for selected response times. The horizontal axis shows the implementation, and the vertical axis shows the benchmark time. Maintaining continuity with Section \ref{sec:apx}, the thirteen implementations on the left are options available in the \texttt{dfddm()} function, with three indicators to show: the timescale of the approximation (``\textbf{l}arge-time'', ``\textbf{s}mall-time'', or ``\textbf{c}ombined-time''), the paper from which the approximation was taken, and the ``small-time'' summation style if necessary ($S_{\textbf{14}}$ or $S_{\textbf{17}}$). The plot on the left is a zoomed in version of the plot on the right in order to show the more subtle differences between the benchmark data from the implementations available in \texttt{dfddm()}. The response times used as inputs to the implementations were between $0$ and $2$ seconds and input to the density function as a vector; the full parameter space can be found in Table \ref{tbl:parspace-2}. For more details on the graphical elements, see Figure \ref{fig:ben-mtl-vec-bm}.}
	\label{fig:ben-vec}
\end{figure}

Figure \ref{fig:ben-vec} shows the distribution of the benchmark times for each implementation. Immediately we notice that the three existing implementations from the literature are significantly slower than the impementations in \texttt{dfddm()}. Of the implementations in \texttt{dfddm()}, there are the three timescales to discuss. First, there is only one ``large-time'' implementation ($f_\ell$), and that appears to be quite slow. Second, there are three ``small-time'' implementations ($f_s$), and our implementation of the SWSE approximation outperforms the others. Third, there are three ``combined-time'' implementations ($f_c$), and again the combination of the SWSE approximation with the ``large-time'' approximation provided by \textcite{navarro2009fast} outperforms the others. Overall, the ``small-time'' implementations appear faster than the ``combined-time'' ones because all of the response times input were under two seconds. Further analysis of this feature will be discussed imminently.

\begin{figure}
	\begin{center}
		\includegraphics[width=\textwidth]{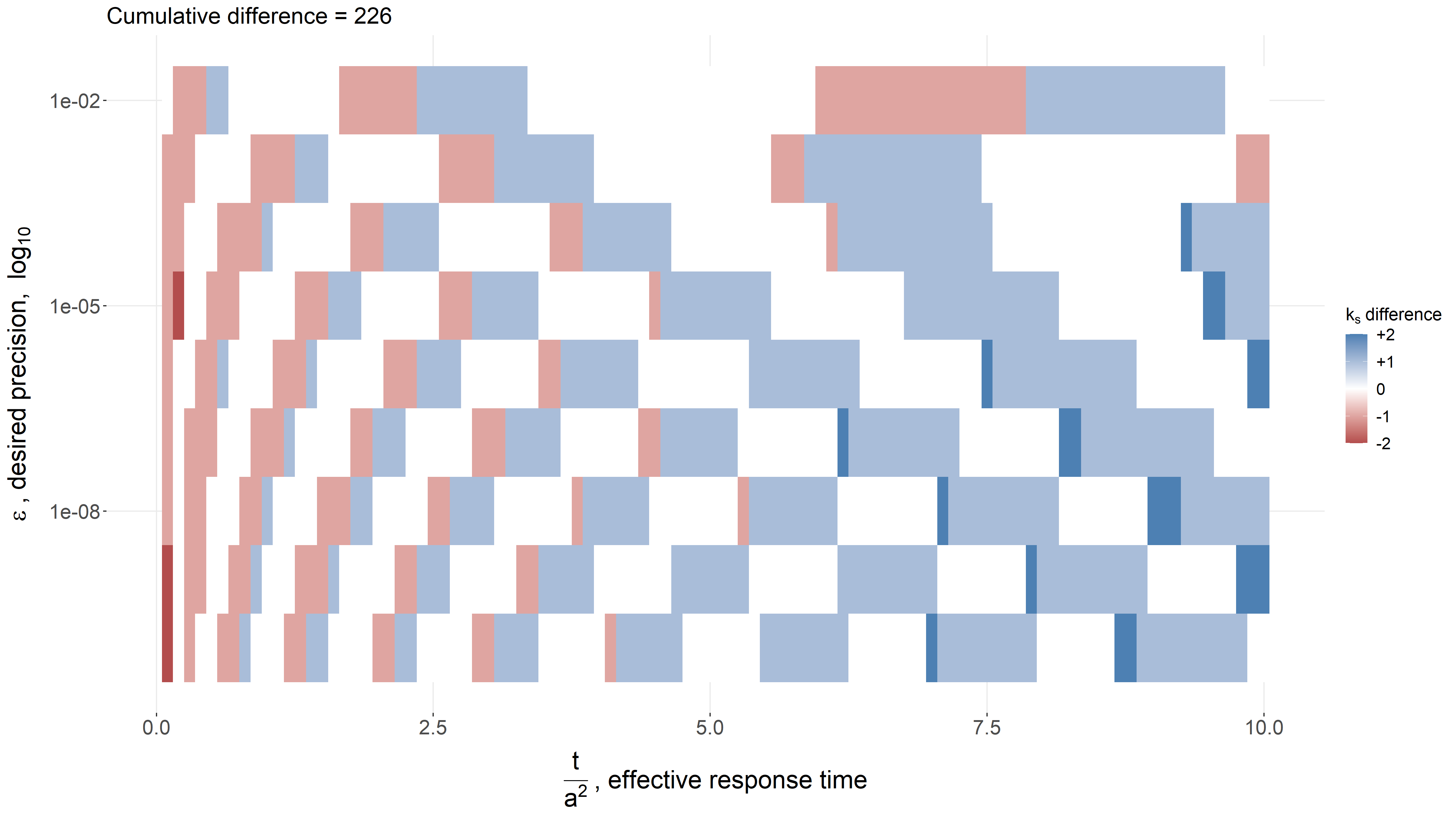}
	\end{center}
	\caption{Difference in number of terms for existing ``small-time'' approximations. The heatmap shows the differences in $k_s^\text{Gon}$ and $k_s^\text{Nav}$ across varying response times and error tolerances. Positive (blue) values indicate that $k_s^\text{Nav}$ uses fewer terms, and negative (red) values indicate that $k_s^\text{Gon}$ uses fewer terms. Note that $k_s^\text{Gon}$ uses the DDM parameter $w$ in its calculation, but $k_s^\text{Nav}$ does not. In this plot, $w$ is fixed to $0.5$; for additional plots with other values of $w$, see the Supplemental Materials document.}
	\label{fig:ben-ks}
\end{figure}

One surprising result in Figure \ref{fig:ben-vec} is that our implementation of the ``small-time'' approximation provided by \textcite{navarro2009fast} is slightly faster than our implementation of the more recent ``small-time'' approximation proposed by \textcite{gondan2014even}. This difference in speed is likely because the number of terms required by the \textcite{navarro2009fast} approximation is sometimes smaller than that of the \textcite{gondan2014even} approximation. As shown in Figure \ref{fig:ben-ks}, $k_s^\text{Nav}$ uses fewer terms overall; however, $k_s^\text{Gon}$ typically uses fewer terms for smaller response times, which likely explains the minimal difference across the implementations of the ``combined-time'' approximations that use these two approximation methods. We also considered that the calculation of $k_s^\text{Nav}$ could be faster than that of $k_s^\text{Gon}$ (compare Equation \eqref{eq:ks_gon} with Equation \eqref{eq:ks_nav}). However, we got contradictory benchmark data for the execution times of these two calculations from different computers. Please see the Supplementary Materials document on GitHub for more information about this topic.

To investigate why the ``small-time'' implementations appear to outperform the ``combined-time'' implementations, we perform another round of benchmark testing. For these tests, we will input the response times individually across the larger parameter space, found in Table \ref{tbl:parspace-30}; this will allow us to see the speed of each implementation as the response time varies across the parameter space. For this set of benchmark results only, we will not consider the response time by itself, but rather we will use the \textit{effective response time}, $\hat{t}$. Since the parameters $t$ and $a$ always exist within the infinite sums as a quotient (regardless of timescale), we consider the quantity $\hat{t} = \frac{t}{a^2}$ as a proxy for the response time.

As we do not wish to clutter the results, we consider only six implementations for this testing that are representative of the currently available implementations. We include all three of the currently available implementations from the literature: the implementations from the \texttt{rtdists} and \texttt{RWiener} packages, as well as the implementation provided by \textcite{gondan2014even} that has been included in \texttt{dfddm()}. In addition to the currently available implementations, we also include the fastest implementation of each timescale in \texttt{dfddm()}: our implementation of the ``large-time'' approximation given by \textcite{navarro2009fast}, the SWSE ``small-time'' implementation, and our ``combined-time'' implementation that uses both the SWSE ``small-time'' approximation and the ``large-time'' approximation provided by \textcite{navarro2009fast}.

\begin{figure}
	\begin{center}
		\includegraphics[width=\textwidth]{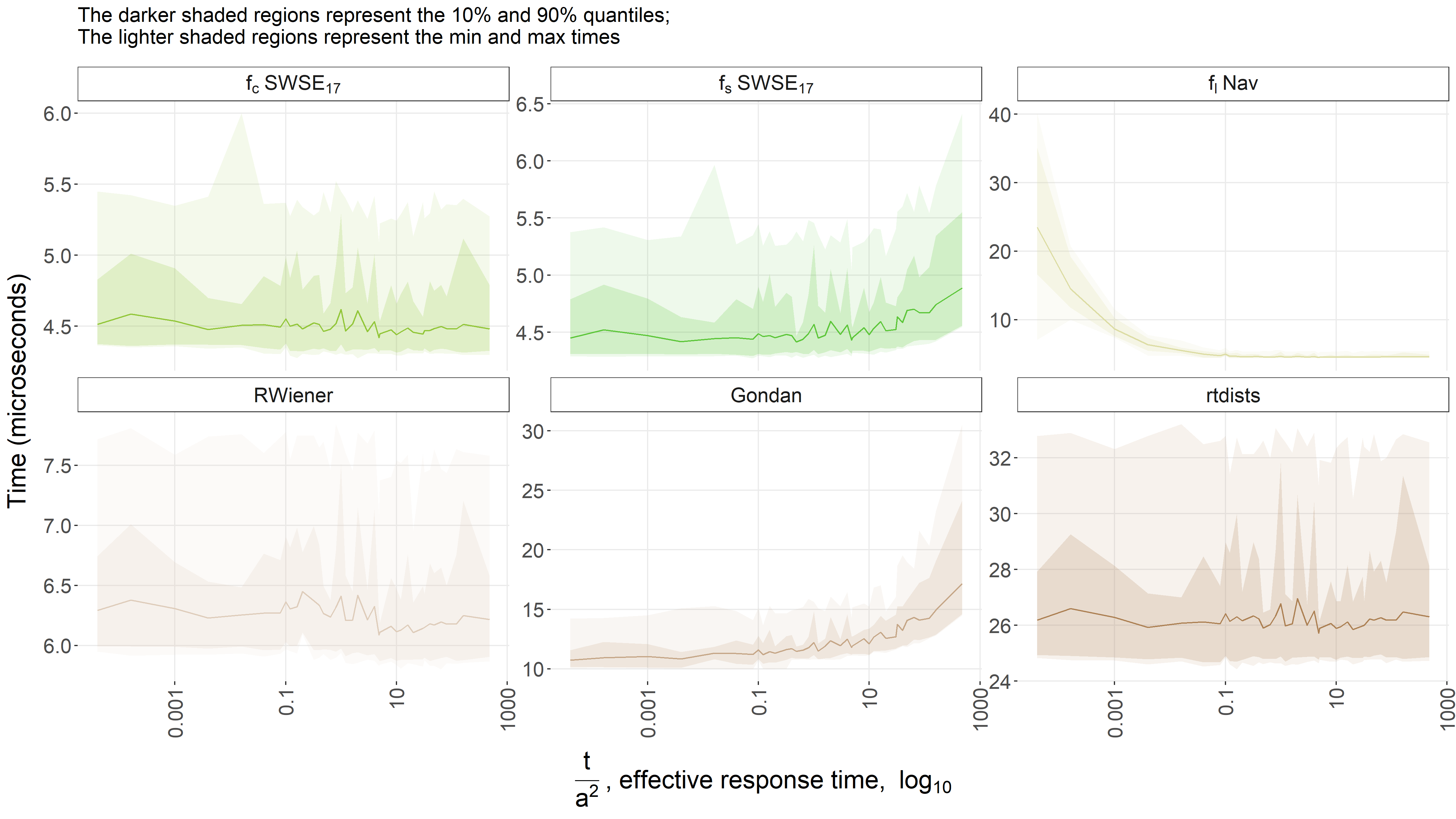}
	\end{center}
	\caption{Effect of effective response time on benchmark times for selected implementations. On the horizontal axis is the effective response time, and the vertical axis displays the benchmark time; note that the vertical axis is not fixed across panels. The dark line in each panel indicates the mean benchmark time; the darker shaded region in each panel shows the 10\% and 90\% quantiles; and the lightly shaded region shows the minimum and maximum benchmark times. The parameter space used in this plot can be found in Table \ref{tbl:parspace-30}.}
	\label{fig:ben-meq}
\end{figure}

Figure \ref{fig:ben-meq} shows that the preconceived notion of the two timescales is accurate. The ``small-time'' implementations perform much better for smaller effective response times and struggle when handling larger effective response times. For the ``large-time'' implementation, the opposite is true; it struggles greatly for smaller effective response times yet is very efficient for larger effective response times. We plotted the benchmark results in the same style for varying the other DDM parameters ($w$, $v$, and $\eta$), but those results were uninteresting and are included in the Supplemental Materials document on GitHub for completeness.

\subsubsection{Data Fitting} \label{sec:ben-imp-fit}

Second, we test the suitable implementations in a simple data fitting routine to estimate the commonly used parameters $a$, $v$, $w$, $t_0$, and $\eta$. While we can use every implementation in \texttt{dfddm()}, we cannot use the implementation from the \texttt{RWiener} \texttt{R} package nor the \texttt{R} implementation from \textcite{gondan2014even} because neither of these two implementations includes $\eta$ in their estimation of the DDM density function.\footnote{While we could use the conversion multiplier $M$ to convert from the constant drift rate density function to the variable drift rate density function, this often leads to inaccurate density estimates. See the Validity Vignette in the \texttt{fddm} CRAN repository at \url{https://cran.r-project.org/package=fddm/vignettes/validity.html\#den-ke} for more information.} Thus we test all of the implementations in \texttt{dfddm()} and the implementation in \texttt{rtdists}.

\begin{figure}[tb!]
	\begin{center}
		\includegraphics[width=\textwidth]{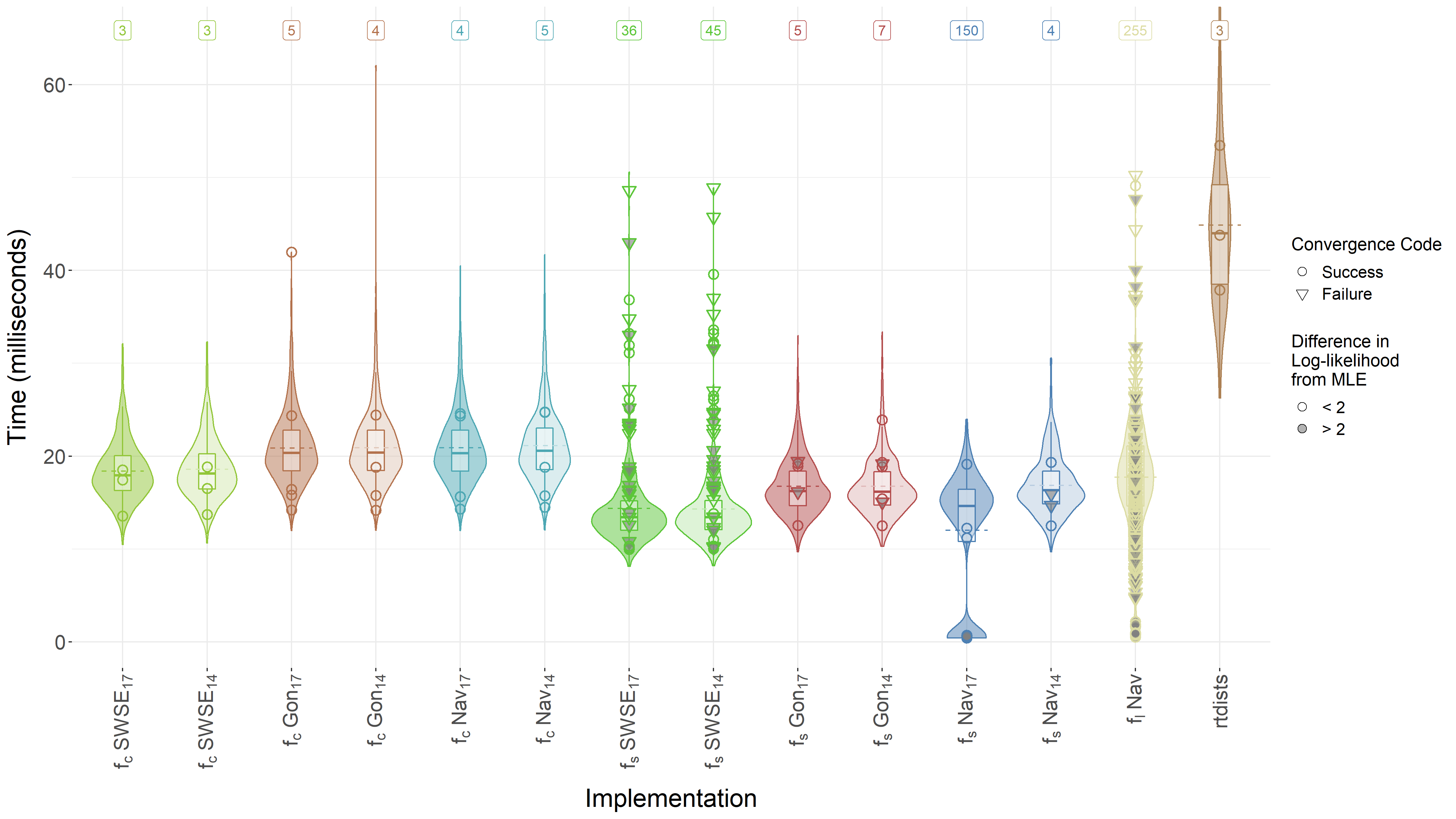}
	\end{center}
	\caption{Benchmark times of different density function approximation implementations for data fitting. The horizontal axis shows the implementation underlying the optimization, and the vertical axis shows the benchmark time. Each distribution shows results from 407 fitting runs (the 37 participants of \nptextcite{trueblood2018impact} times 11 fixed sets of starting values). More details on the data fitting process are included in Section \ref{sec:ben-met-fit}, and more details on the graphical elements are given in Figure \ref{fig:ben-mtl-fit-bm}.}
	\label{fig:ben-fit-bm}
\end{figure}

Figure \ref{fig:ben-fit-bm} shows the benchmark times and notable convergence issues in the data fitting. Similarly to the results shown in Figure \ref{fig:ben-vec}, the implementations in \texttt{dfddm()} are faster than the currently available implementation in \texttt{rtdists}. However, the magnitude of the difference is a lot smaller; whereas in Figure \ref{fig:ben-vec} the benchmark time for the \texttt{rtdists} implementation is about five times that of the available options in \texttt{dfddm()}, the benchmark time for the \texttt{rtdists} implementation is only about twice that of the available options in \texttt{dfddm()}.

Of the implementations in \texttt{dfddm()}, there are again three timescales to consider. First, the ``large-time'' implementation alone appears quite erratic by showing many unsuccessful optimization attempts. Second, the SWSE ``small-time'' implementation appears to be slightly faster than that of \textcite{gondan2014even} or \textcite{navarro2009fast}, but it is comparatively more problematic with many unsuccessful optimization attempts. Third, the combination of the SWSE ``small-time'' approximation and the ``large-time'' approximation from \textcite{navarro2009fast} is the fastest and most stable implementation for data fitting, outperforming the other implementations both in terms of benchmark time and successful optimizations.

\begin{figure}[tb!]
	\begin{center}
		\includegraphics[width=\textwidth]{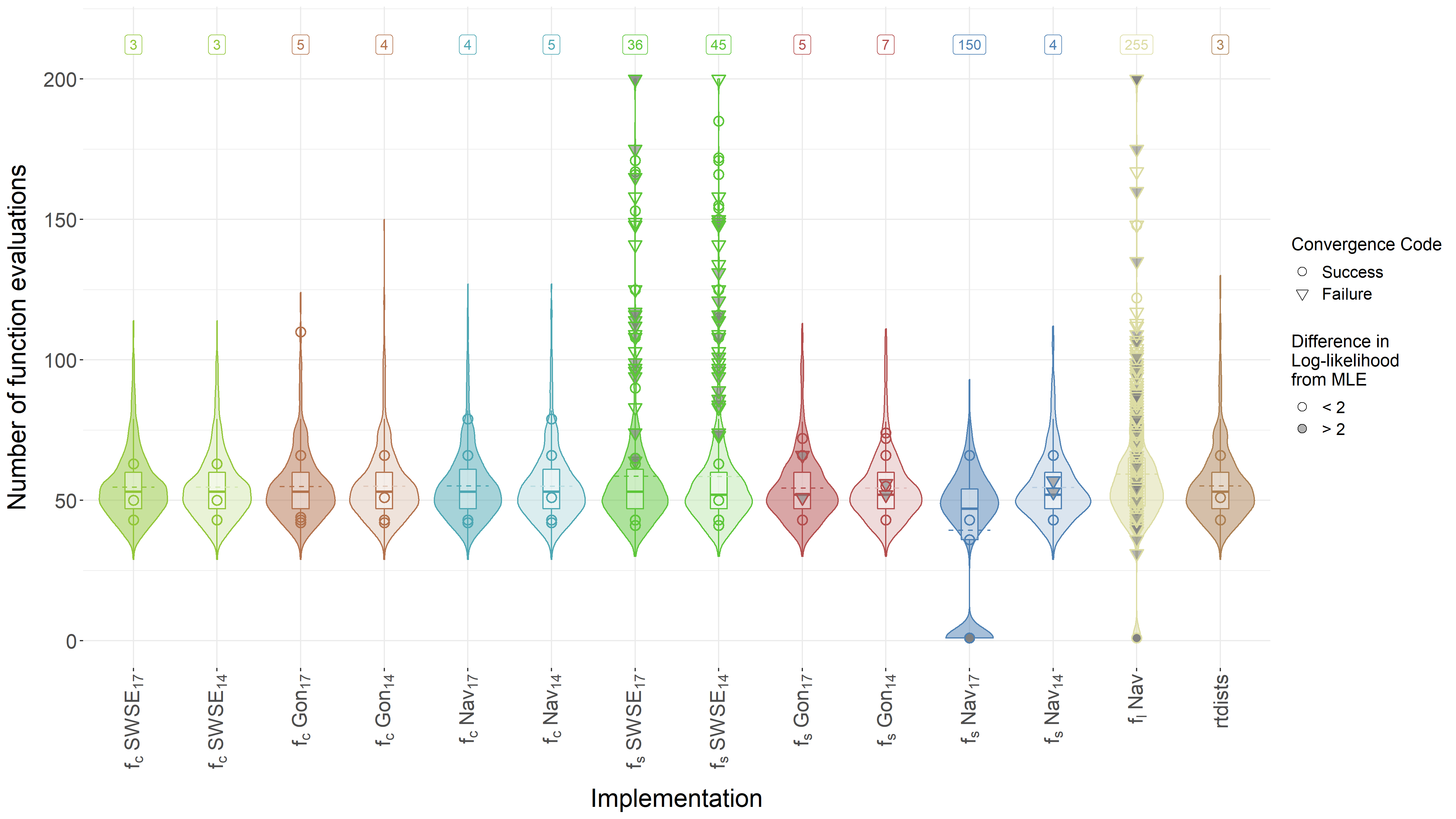}
	\end{center}
	\caption{Number of calls to the likelihood function for different density function approximation implementations during data fitting. The horizontal axis shows the implementation underlying the optimization, and the vertical axis shows the number of likelihood evaluations. Each distribution shows results from 407 fitting runs (the 37 participants of \nptextcite{trueblood2018impact} times 11 fixed sets of starting values). More details on the data fitting process are included in Section \ref{sec:ben-met-fit}, and more details on the graphical elements are given in Figure \ref{fig:ben-mtl-fit-bm}.}
	\label{fig:ben-fit-fe}
\end{figure}

In addition to the raw computation speed of each implementation, we also recorded the number of function evaluations that the optimization process had to make in order to find the optimum parameter estimates. In general, fewer function calls suggest a more simply shaped parameter space, making it easier for the optimization routine to quickly and accurately determine the optimum parameter estimates. In Figure \ref{fig:ben-fit-fe}, we see that most of the implementations used required approximately the same number of function calls. The SWSE ``small-time'' implementation and our implementation of the ``large-time'' approximation given by \textcite{navarro2009fast} all showed issues in benchmark time as well as function evaluations; however, these two implementations are the most effective when combined.

\section{Discussion} \label{sec:dis}

We performed benchmark tests to determine the fastest approximation method for the density function of the DDM. In addition to directly comparing the existing approximation methods for speed, we also implemented a novel approximation method and compared it to the existing ones. There are three main results to discuss.

First, we implemented two different styles of summation for the ``small-time'' algorithms, $S_{14}$ and $S_{17}$. There is not much to distinguish these two summation styles in the benchmark results; however, we note that the $S_{17}$ style appears to marginally outperform the $S_{14}$ style for most implementations. Whereas the relative performance of these two ``small-time'' summation styles may vary slightly depending on the implementation and computer used for the calculation, the $S_{17}$ style uses at most as many terms in the truncated summation as the $S_{14}$ style (see Section \ref{sec:apx-sma} for details). For these reasons, the $S_{17}$ style is the recommended and default ``small-time'' summation style for the \texttt{fddm} package.

Second, a ``combined-time'' approximation is better to use in a general setting than either a pure ``large-time'' or a pure ``small-time'' approximation. It handles both the small and large effective response times that arise during a standard optimization routine, and this is useful for two reasons. First, it eliminates the inefficiencies of a pure ``small-time'' approximation when encountering large effective response times, and vice-versa for a pure ``large-time'' approximation. Combining these two approximations yields an overall more efficient algorithm with benchmark results tightly clustered around the median. Second, it eliminates the inaccuracies and convergence issues that arise in a data fitting scenario from evaluating the PDF for small or large effective response times using a pure ``large-time'' or ``small-time'' approximation, respectively. As data fitting is a common use for evaluating the DDM density function, this increased stability is paramount to any implementation.

Lastly, our implementations of these density function approximations are noticeably faster than the current standards found in \texttt{R} packages (i.e., \texttt{rtdists} and \texttt{RWiener}) and unpackaged code from the literature (i.e., from \textcite{gondan2014even}). In particular, our implementation of our novel ``combined-time'' approximation method is the fastest currently available way to reliably approximate the DDM density functions. Most of the increase in speed for this implementation comes from the use of the faster \texttt{C++} programming language. However, the SWSE ``small-time'' approximation method further increased the speed, and using the ``large-time'' approximation method from \textcite{navarro2009fast} removed the outliers in the tail of the benchmark results.

Although our implementation of our novel approximation method is faster than the currently available implementations in terms of execution time for a set of fixed, selected response times, the benchmark times across the implementations are more similar for data fitting (Figure \ref{fig:ben-fit-bm} compared to Figure \ref{fig:ben-vec}). This decreased difference in the benchmark results is due to the overhead of running an optimization routine in \texttt{R}. Despite its impressive vectorization, \texttt{R} has a lot of overhead when calling functions. Since an optimization routine will make many function calls (e.g., to the likelihood function), the overhead of these function calls makes up a significant part of the overall benchmark time for each approximation. If speed is of the utmost importance, we recommend avoiding \texttt{R} in favor of a programming language that has less overhead in function calls (e.g., pure \texttt{C++}).

Overall, our results show that the implementation that combines the SWSE ``small-time'' approximation method with the ``large-time'' approximation method from \textcite{navarro2009fast} is both the fastest and most consistent across broad parameter spaces. If the density function is \textit{only} being evaluated for \textit{small} effective response times, then the implementation of the SWSE ``small-time'' approximation method may be the fastest option. However, since a very common use for evaluating the DDM density function is data fitting, an optimization function would likely encounter convergence issues when exploring parameter values because the pure ``small-time'' approximation methods, especially the SWSE method, have shown to be problematic when being evaluated at large effective response times. In sum, we suggest researchers interested in fitting the DDM use the default implementation from the \texttt{fddm} package for the most robust and efficient approximation to the DDM density function.

\printbibliography

\newpage
\begin{appendices}
\setcounter{equation}{0}
\renewcommand{\theequation}{\Alph{section}.\arabic{equation}}

\section{Mathematical Proofs} \label{app:mat}

This section will present the various mathematical proofs that we have referenced throughout this paper.

\subsection{Derivation of the ``Large-Time'' Density Function with Variable Drift Rate} \label{app:mat-lar}
In Section \ref{sec:apx} \nameref{sec:apx}, we provided a version of the ``large-time'' density function that allowed the drift rate to vary across trials; we will present the steps of this derivation here. We begin by writing the ``small-time'' density function that allows for inter-trial variability, Equation \eqref{eq:den-var-s} from Section \ref{sec:apx}:
\begin{equation}
	f_\ell(t | v, \eta^2, a, w) = \frac{a}{\sqrt{2 \pi t^3 \left( 1 + \eta^2 t \right)}}
	e^{ \left( \frac{\eta^2 a^2 w^2 -2vaw -v^2 t}{2 (1 + \eta^2 t)} \right)}
	\sum_{j = -\infty}^{\infty} (w + 2j) e^{ \left( -\frac{a^2}{2t} \left( w + 2j \right)^2 \right)}.
\end{equation}

Next, we manipulate the infinite sum into the form given by \textcite{navarro2009fast}. Note that in their original paper, \textcite{navarro2009fast} scale $t$ prior to including it in the sum. They use $t := \frac{t}{a^2}$, which explains why we have additional $a^2$ terms compared to their formulation. This manipulation gives
\begin{equation}
	f_\ell(t | v, \eta^2, a, w) = \frac{1}{a^2 \sqrt{1 + \eta^2 t}}
	e^{ \left( \frac{\eta^2 a^2 w^2 -2vaw -v^2 t}{2 (1 + \eta^2 t)} \right)}
	\sum_{j = -\infty}^{\infty} \frac{a^3}{\sqrt{2 \pi t^3}} (w + 2j) e^{ \left( -\frac{a^2}{2t} \left( w + 2j \right)^2 \right)}.
\end{equation}

Since the ``large-time'' and ``small-time'' density functions are equivalent in their constant drift rate versions, we can equate them. The multiplicative term outside the infinite sum is the same in both functions, thus their infinite sums must be equal as well. This equivalence lets us substitute the infinite sum from the expression above with the infinite sum from the ``large-time'' density function from \textcite{navarro2009fast}:
\begin{equation}
	f_\ell(t | v, \eta^2, a, w) = \frac{1}{a^2 \sqrt{1 + \eta^2 t}}
	e^{ \left( \frac{\eta^2 a^2 w^2 -2vaw -v^2 t}{2 (1 + \eta^2 t)} \right)}
	\sum_{j = 1}^{\infty} \pi j \sin \left( j w \pi \right) e^{ \left( -\frac{j^2 \pi^2 t}{2a^2} \right)}.
\end{equation}

Removing the $\pi$ from inside of the infinite sum yields the final form of the ``large-time'' density function, as described in Equation \eqref{eq:den-var-l} from Section \ref{sec:apx}:
\begin{equation}
	f_\ell(t | v, \eta^2, a, w) = \frac{\pi}{a^2 \sqrt{1 + \eta^2 t}}
	e^{ \left( \frac{\eta^2 a^2 w^2 -2vaw -v^2 t}{2 (1 + \eta^2 t)} \right)}
	\sum_{j = 1}^{\infty} j \sin \left( j w \pi \right) e^{ \left( -\frac{j^2 \pi^2 t}{2a^2} \right)}.
\end{equation}

\subsection{Equivalence of ``Small-Time'' Summation Styles} \label{app:mat-sum}
We will demonstrate that the two seemingly different summation styles from Section \ref{sec:apx} \nameref{sec:apx} actually produce the same series. The two styles are
\begin{equation} \label{app-eq:sum}
	\begin{aligned}
		S_{14} =& \sum_{i = -\infty}^{\infty} (w + 2i) e^{ \left( -\frac{a^2}{2t} \left( w + 2i \right)^2 \right)},\\[2ex]
		S_{17} =& \sum_{j = 0}^{\infty} \left( -1 \right)^j r_j e^{ \left( -\frac{a^2 r_j^2}{2t} \right)},\\
		&r_j = \begin{cases}
			j + w & \text{ if $j$ is even,}\\
			j + 1 - w & \text{ if $j$ is odd.}
		\end{cases}\\
	\end{aligned}
\end{equation}

Let the $S_{14}$ style terms be denoted as $\sett{b_i}_{-\infty}^\infty$. Similarly, let the $S_{17}$ style terms be denoted as $\sett{c_j}_0^\infty$. Instead of the usual ordering, consider the $S_{14}$ style terms as the series $\sett{b_0, b_{-1}, b_1, \dots, b_{-i}, b_i, b_{-(i+1)}, b_{i+1}, \dots}$; we denote this ordering by $\sett{b}_0^{\pm\infty}$. Our goal is to show that this reordering of the $S_{14}$ style terms produces exactly the same series as the natural ordering of the $S_{17}$ style terms.

We need to show that the sets $\sett{(-1)^j r_j}_{j=0}^\infty$ and $\sett{w + 2i}_{i = 0}^{\pm\infty}$ contain precisely the same elements in the same order. We also need to show that the sets $\sett{\frac{a^2}{2t} (r_j)^2}_{j=0}^\infty$ and $\sett{\frac{a^2}{2t} (w + 2i)^2}_{i = 0}^{\pm\infty}$ also contain precisely the same elements in the same order, but we will handle this later.

First, consider even values of $j$ in $\sett{(-1)^j r_j}_{j=0}^\infty$. Notice that $(-1)^j r_j = w + j$ in this case. Since $j$ is even, we can write $j = 2n$, for $n \in \Nat$. Thus we have $(-1)^j r_j = w + 2n$, and the value of each $S_{17}$ style term with even $j$ has precisely one corresponding term of equal value in the $S_{14}$ style with nonnegative index $i$. As the mapping $j \mapsto 2n$ is order-preserving (i.e., monotone), the order-preserving identity map $n \mapsto i$ completes the bijection from the $S_{17}$ style terms with even index $j$ to the $S_{14}$ style terms with nonnegative index $i$.

Next, consider odd values of $j$ in $\sett{(-1)^j r_j}_{j=0}^\infty$. Notice that $(-1)^j r_j = w - (j+1)$ in this case. Since $j$ is odd, $j+1$ is even; then we can write $j+1 = 2m$, for $m \in \Nat$. Thus we have $(-1)^j r_j = w - 2m$, and the value of each $S_{17}$ style term with odd $j$ has precisely one corresponding term of equal value in the $S_{14}$ style with negative index $i$. As the mapping $j+1 \mapsto 2m$ (or equivalently, $j \mapsto 2m-1$) is order-preserving, the order-preserving identity map $m \mapsto i$ completes the bijection from the $S_{17}$ style terms with odd index $j$ to the $S_{14}$ style terms with negative index $i$.

Now we will prove the equality of the squared terms in the exponentials. Notice that
\begin{equation}
	\sett{ r_j^2 }_{j=0}^\infty = \sett{ (-1)^{2j} r_j^2 }_{j=0}^\infty = \sett{ \left( (-1)^j r_j \right)^2 }_{j=0}^\infty.
\end{equation}

Since we have just shown that the sets $\sett{(-1)^j r_j}_{j=0}^\infty$ and $\sett{w + 2i}_{i = 0}^{\pm\infty}$ are equivalent, we can index both styles of terms with the set of natural numbers (e.g., use the $S_{17}$ style indexing). Then it follows that the sets $\sett{ \frac{a^2}{2t} r_j^2 }_{j=0}^\infty$ and $\sett{ \frac{a^2}{2t} \left( w + 2i \right)^2 }_{i = 0}^{\pm\infty}$ are also equivalent because (1) multiplying by a constant (i.e., $\frac{a^2}{2t}$), (2) the square function, and (3) the exponential function are all order-preserving bijections for all nonnegative real numbers.

Since both styles of summation produce precisely the same terms, they are equivalent given the ordering defined above (i.e., $\sett{b}_0^{\pm\infty} = \sett{c_j}_0^\infty$).

\subsection{Summations as Convergent Decreasing Series} \label{app:mat-inf}

This section will prove that both the ``large-time'' and ``small-time'' infinite summations can be treated as convergent decreasing series.

\subsubsection{Large-Time} \label{app:mat-inf-lar}

In Section \ref{sec:apx-lar}, we mentioned that the infinite sum in the ``large-time'' density functions, Equations \ref{eq:den-con-l} and \ref{eq:den-var-l}, converges. Note that the infinite sum is identical in both the constant drift rate and variable drift rate variants of the density function, so this argument applies to both Equation \ref{eq:den-con-l} and Equation \ref{eq:den-var-l}.

The sum in question is
\begin{equation}
	\sum_{j = 1}^{\infty} j \sin \left( j w \pi \right) e^{ \left( -\frac{j^2 \pi^2 t}{2a^2} \right)}.
\end{equation}

We will prove that the sum is absolutely convergent and, thus, convergent. Upon taking the absolute value inside the sum, we note that $j$ and the exponential are unaffected as they are strictly positive anyway. Since $\abs{\sin(x)} \leq 1 ~ \forall ~ x \in \Real$, we can eliminate the $\sin$ term and write the absolute sum:
\begin{equation}
	\sum_{j = 1}^{\infty} \abs{j \sin \left( j w \pi \right) e^{ \left( -\frac{j^2 \pi^2 t}{2a^2} \right)}} \leq \sum_{j = 1}^{\infty} j e^{ \left( -\frac{j^2 \pi^2 t}{2a^2} \right)}.
\end{equation}

We will proceed by using the integral test for convergence. Let $f(j) = j e^{ \left( -\frac{j^2 \pi^2 t}{2a^2} \right)}$, a nonnegative function. Note that $f'(j) = e^{ \left( -\frac{j^2 \pi^2 t}{2a^2} \right)} \left( 1 - j^2 \frac{\pi^2 t}{a^2} \right)$, so $f$ is monotonically decreasing for $j > \frac{a}{\pi \sqrt{t}}$. For practical purposes\footnote{As shown in this paper, the ``large-time'' approximation performs poorly for small response times. When combined with the SWSE ``small-time'' approximation, the usual transition point from using the ``small-time'' to the ``large-time'' approximation occurs on the scale of $0.5$ seconds.}, we would expect $a \leq 2$ and $t \geq 0.5$; substituting these values gives that $f$ monotonically decreases for $j > 0.9$, which is satisfactory for our needs.

Thus, we can proceed with the integral test:
\begin{equation}
	\int_1^\infty j e^{ \left( -\frac{j^2 \pi^2 t}{2a^2} \right)} ~dj = \lim\limits_{b \to \infty} \int_1^b e^{ \left( -\frac{j^2 \pi^2 t}{2a^2} \right)} j ~dj.
\end{equation}

Using the substitution $u = \frac{j^2 \pi^2 t}{2a^2}$ (and so $du = \frac{\pi^2 t}{a^2} j ~dj$ and $b \mapsto b' = \frac{b^2 \pi^2 t}{2a^2}$), we have
\begin{equation}
	\begin{aligned}
		\lim\limits_{b' \to \infty} \int_1^{b'} \frac{a^2}{\pi^2 t} e^{-u} ~du &= \frac{a^2}{\pi^2 t} \lim\limits_{b' \to \infty} \left[ -e^{-u} \right]_1^{b'}\\
		&= \frac{a^2}{\pi^2 t} \lim\limits_{b' \to \infty} \left[ -e^{-b'} + e^{-1} \right]\\
		&= \frac{a^2}{\pi^2 t} \cdot \frac{1}{e},
	\end{aligned}
\end{equation}
which is finite. Thus, the absolute sum converges and so does the original sum.

\subsubsection{Small-Time} \label{app:mat-inf-sma}

In Section \ref{sec:apx}, we reintroduced a method for truncating the infinite sum from Equation \eqref{eq:den-con-s}; the idea was essentially to stop adding terms to the truncated sum once the terms were smaller than the allowed error tolerance, $\epsilon$. Note that for practical purposes of evaluating the infinite sum, we rescale the given error tolerance by the reciprocal of the multiplicative term in front of the infinite sum: $\epsilon' = \frac{1}{a} \sqrt{2 \pi t^3} \exp \left( vaw + \frac{v^2 t}{2} \right) \epsilon$. The legitimacy of this algorithm hinges on the alternating series test\footnote{The alternating series test is also known as Leibniz's rule, Leibniz's test, or the Leibniz criterion.}, which places an upper bound on the truncation error of this sum. Thus to prove the validity of this method, we must show that the terms in the summation (1) alternate, (2) monotonically decrease in absolute value, and (3) approach zero in the limit. Since we just demonstrated that the two summation styles ($S_{14}$ and $S_{17}$) are equivalent, we will only present the full argument for the $S_{14}$ style summation as defined below:
\begin{equation} \label{app-eq:s14}
	S_{14} = \sum_{j = -\infty}^{\infty} (w + 2j) e^{ \left( -\frac{a^2}{2t} \left( w + 2j \right)^2 \right)}.
\end{equation}

Let $b_j = (w + 2j) \exp \left( -\frac{a^2}{2t} \left( w + 2j \right)^2 \right)$. Consider the terms of the sum in the order $\sett{b_0, b_{-1}, b_1, \dots, b_{-j}, b_j, b_{-(j+1)}, b_{j+1}, \dots}$.

First we show that the sign of consecutive terms alternates. Notice that the sign of each term only depends on $w + 2j$, and recall that $w \in (0,1)$. The $j = 0$ term is positive, and then the rest of the terms alternate signs because $\abs{2j} > w ~ \forall ~ j \in \sett{1, 2, \dots}$. It remains to show that $\abs{b_j}$ decreases monotonically and approaches zero in the limit as $j \to \infty$.

Next we show that the absolute value of the series decreases monotonically after a certain number of terms. For large values of $j$, the exponential term will dominate the linear term. However, it is important to note that for some values of $t$, $a$, and $w$ the value of the terms in the summation may actually increase in absolute value before decreasing. To find this critical point we define $J_{14}$ as the smallest index in the series such that all terms after the $J_{14}\th$ are monotonically decreasing in absolute value according to the ordering from the previous paragraph. Then we have
\begin{equation}
	\begin{aligned}
		\sum_{j = -\infty}^{\infty} (w + 2j) e^{ \left( -\frac{a^2}{2t} \left( w + 2j \right)^2 \right)} &=
		\sum_{j = -J_{14}}^{J_{14}} (w + 2j) e^{ \left( -\frac{a^2}{2t} \left( w + 2j \right)^2 \right)}\\
		& \hspace{3mm} + \sum_{j = -(J_{14}+1)}^{-\infty} (w + 2j) e^{ \left( -\frac{a^2}{2t} \left( w + 2j \right)^2 \right)}\\
		& \hspace{3mm} + \sum_{j = J_{14}+1}^{\infty} (w + 2j) e^{ \left( -\frac{a^2}{2t} \left( w + 2j \right)^2 \right)},
	\end{aligned}
\end{equation}
where both of the infinite sums on the right side of the equation (lines two and three) will combine to satisfy the conditions of the alternating series test. It remains to calculate the value of $J_{14}$ and provide conditions for its existence.

We treat the individual terms of the summation as a function of $j$, then take the derivative of this function with respect to $j$ in order to determine the index at which the terms shift from increasing to decreasing. Define $g_{14}(j) = (w + 2j) \exp \left( -\frac{a^2}{2t} \left( w + 2j \right)^2 \right)$, for $j \in \Int$. Then using the chain rule and setting equal to zero, we have
\begin{equation}
	g_{14}'(j) = \left( 2 -\frac{2a^2}{t} (w + 2j)^2 \right) e^{ \left( -\frac{a^2}{2t} \left( w + 2j \right)^2 \right)} = 0,
\end{equation}
which implies that
\begin{equation}
	(w + 2j)^2 = \frac{t}{a^2}.
\end{equation}
Solving for $j$ and giving it the label $j_{14}^*$ gives
\begin{equation}
	j_{14}^* = \frac{\sqrt{t}}{2a} - \frac{w}{2}.
\end{equation}

Thus, the first term whose index is at least $j_{14}^*$ is the first term of the absolutely and monotonically decreasing series. Note that $\frac{\sqrt{t}}{2a} - \frac{w}{2} > -0.5$ because $t \geq 0$, $a > 0$, and $w \in (0, 1)$; we have two cases to discuss.

First, if $-0.5 < j_{14}^* < 0$, then the term in the series with index $-1$ is the first term of the absolutely and monotonically decreasing series. In this case, we must only include the $0^\text{th}$ term of the series before applying the alternating series test.

Second, if $j_{14}^* \geq 0$, then we must include terms in the series whose indices are at most $j^*$ in absolute value; that is, the indices $\sett{-\floor{j_{14}^*}, \dots, 0, \dots, \floor{j_{14}^*}}$. Then, the term in the series with index $-(\floor{j_{14}^*} + 1) = -\ceil{j_{14}^*}$ is the first term of the absolutely and monotonically decreasing series\footnote{As mentioned in the main body of this paper, $\floor{\cdot}$ denotes the floor function: the argument is rounded down to the nearest integer that is less than or equal to the argument. Similarly, $\ceil{\cdot}$ denotes the ceiling function: the argument is rounded up to the nearest integer that is greater than or equal to the argument.}. As the terms with negative indices are greater in absolute value than the corresponding terms with positive indices (e.g., $\abs{b_{-j}} > \abs{b_j}$), it follows that the term with the greatest absolute value has a negative index. Thus, the absolutely and monotonically decreasing series should start with the term whose index is $-\ceil{j_{14}^*}$.

Therefore, we set $J_{14}$ -- the smallest index in the $S_{14}$ series such that all terms after the $J_{14}\th$ are monotonically decreasing in absolute value -- as
\begin{equation}
	J_{14} = \max \sett{0, ~ \floor{j_{14}^*}} = \max \sett{0, ~ \floor{\frac{\sqrt{t}}{2a} - \frac{w}{2}}}.
\end{equation}

Note that there exists a bijective correspondence between the terms of the $S_{14}$ style series and the $S_{17}$ style series. Using this relationship, we note that the non-negative indices in the $S_{14}$ style correspond to the even indices in the $S_{17}$ style. Thus, we can convert $J_{14}$ into $J_{17}$ by the following:
\begin{equation}
	J_{17} = \max \sett{0, ~ 2 \cdot \floor{j_{14}^*}} = \max \sett{0, ~ \floor{\frac{\sqrt{t}}{a} - w}}.
\end{equation}

As an extra precaution, we can take the second derivative of $g_{14}(j)$ to confirm the concavity of $g_{14}(j)$ and show that $j_{14}^*$ is indeed a maximum and not a minimum. Again using the chain rule to take the second derivative of $g_{14}(j)$ with respect to $j$ yields upon simplification
\begin{equation}
	g_{14}''(j) = \left( \frac{2a^2}{t} (w + 2j)^2 - 6 \right) \frac{2a^2}{t} e^{ \left( -\frac{a^2}{2t} \left( w + 2j \right)^2 \right)}.
\end{equation}

Substituting $j_{14}^* = \frac{\sqrt{t}}{2a} - \frac{w}{2}$ for $j$ in the expression for $g_{14}''(j)$ gives the concavity of the function at the critical point. Note that $\frac{2a^2}{t} e^{ \left( -\frac{a^2}{2t} \left( w + 2j \right)^2 \right)} > 0 ~ \forall ~ j \in \Int$, so we define
\begin{equation}
	C(j) = \frac{2a^2}{t} e^{ \left( -\frac{a^2}{2t} \left( w + 2j \right)^2 \right)}
\end{equation}
for expositional purposes because we only need to consider the positivity of $g_{14}''(j_{14}^*)$ and not its magnitude. Again, note that $C(j) > 0 ~ \forall ~ j \in \Int$. Then we have
\begin{equation}
	g_{14}''\left( \frac{\sqrt{t}}{2a} - \frac{w}{2} \right) = \left( \frac{2a^2}{t} \left(w + 2 \frac{\sqrt{t}}{2a} - \frac{w}{2} \right)^2 - 6 \right) \cdot C\left( j_{14}^* \right) = -4 \cdot C\left( j_{14}^* \right) < 0.
\end{equation}

This demonstrates that $g_{14}(j)$ is concave down at $j_{14}^*$ and thus the sequence of terms $\sett{b_j}$ is monotonically decreasing in absolute value for $j \geq J_{14}$.

Lastly we will show that $\abs{b_j}$ approaches zero in the limit, that is, $\lim\limits_{j \to \infty} \abs{b_j} = 0$. Expanding this gives
\begin{equation}
	\lim_{j \to \infty} \abs{b_j} = \lim_{j \to \infty} \abs{\frac{(w + 2j)}{\exp \left( -\frac{a^2}{2t} \left( w + 2j \right)^2 \right)}}.
\end{equation}

Applying L'Hospital's rule with respect to $j$ yields
\begin{equation}
	\lim_{j \to \infty} \abs{{b_j}} = \lim_{j \to \infty} \frac{2}{\abs{-\frac{2a^2}{t} (w + 2j)^2 \exp \left( -\frac{a^2}{2t} \left( w + 2j \right)^2 \right)}} = 0.
\end{equation}

Since this summation satisfies the conditions of the alternating series test (and by extension the $S_{17}$ style does as well), the truncation error for stopping the summation at the $K\th$ term is at most the absolute value of the next term in the series. We will denote this term $b_{K\pm}$ because it is either $b_{K+1}$ if $K \geq 0$ or $b_{K-1}$ if $K < 0$, for $K \in \Int$. Thus, as long as $b_{K\pm} < \epsilon'$ and $\abs{K} \geq J_{14}$, including at least $K$ terms in our approximation will guarantee that the overall truncation error is less than the provided allowed error tolerance $\epsilon$.

\end{appendices}

\end{document}